\documentclass[preprint,aps,floatfix,showpacs]{revtex4}
\usepackage{graphicx}
\usepackage[usenames]{color}

\newcommand{\beq}{\begin{equation}}
\newcommand{\eeq}{\end{equation}}
\newcommand{\bea}{\begin{eqnarray}}
\newcommand{\eea}{\end{eqnarray}}
\usepackage{lscape}
\begin{document}
\preprint{\vbox{\hbox{ JLAB-THY-11-1405} }}
%\vspace{0.5cm}
\title{\phantom{x}
%\vspace{-0.5cm}    Negative Parity  Baryon Decays
% in the ${\mathbf 1}{\mathbf /}{\rm \mathbf N_{\mathbf c}}$ Expansion}

\vspace{-0.5cm}    Partial Decay Widths of Negative Parity  Baryons
 in the ${\mathbf 1}{\mathbf /}{\rm \mathbf N_{\mathbf c}}$ Expansion}

\author{
C.~Jayalath $^{a,b,c}$
\thanks{e-mail:jayalath@jlab.org},
J.~L.~Goity $^{a,b}$ \thanks{e-mail: goity@jlab.org},
E.~Gonz\'alez de Urreta $^{d,e}$
\thanks{e-mail: emilianogdeurreta@gmail.com},
N.~N.~Scoccola $^{d,e,f}$
\thanks{e-mail: scoccola@tandar.cnea.gov.ar}}

\affiliation{
$^a$Department of Physics, Hampton University, Hampton, VA 23668, USA.\\
%}
%\affiliation{
%
$^b$Thomas Jefferson National Accelerator Facility, Newport News, VA 23606, USA.\\
$^c$Department of Physics, Peradeniya University, Peradeniya 20400,  Sri Lanka.\\
$^d$Department of Theoretical Physics, Comisi\'on Nacional de Energ\'{\i}a
At\'omica, 1429  Buenos Aires, Argentina.\\
$^e$CONICET, Rivadavia 1917,  1033  Buenos Aires, Argentina.\\
$^f$Universidad Favaloro, Sol{\'\i}s 453,  1078 Buenos Aires, Argentina.
}
%\date{\today}
\begin{abstract}
%(Last modified: Jose August 4 )

The partial decay widths of lowest lying negative parity baryons belonging to the {\bf 70}-plet  of $SU(6)$ are analyzed in the framework of the $1/N_c$ expansion.
The channels considered are those with single pseudoscalar meson emission. The analysis is carried out to sub-leading order in
$1/N_c$ and to first order in $SU(3)$ symmetry breaking. Conclusions about the magnitude of $SU(3)$ breaking effects along with predictions for some unknown or poorly determined  partial decay widths of 
known resonances are obtained.
\end{abstract}


\maketitle

\section{Introduction}

The extensive experimental programs at various facilities, in
particular  Jefferson Lab, MAMI, ELSA, GRAAL and BES, where photo-
and electro-production data as well as $J/\psi$ decays  give
unprecedented access to baryon resonance parameters,    will  lead
to significant improvements over  the current  knowledge of
resonance  masses and partial widths.  With this progress,
further sharpening of the theoretical approaches is  needed.
Although successful to  a  remarkable extent, the study of decays
with quark models is affected by numerous choices about the
mechanism of decay \cite{CR-QM}, which result in model
dependencies which are difficult to quantify.  An alternative
approach, which gives a systematic connection to QCD is based  on
the $1/N_c$ expansion.  As formulated for baryons, it allows one
to represent quantities and observables in a systematic expansion
in effective operators \cite{DM,CGKM,Goity,PY} where the
coefficients   encode the unknown dynamics.  The expansion is
ordered in powers of $1/N_c$, and the mentioned coefficients are
determined by fitting to data.  Tests of the feasibility of the
expansion at each order are provided by relations which are
independent of those coefficients, and by  whether the magnitude
of next to leading order effects are or are not of natural size.

The $1/N_c$ expansion for  excited  baryons is based on the classification
of   states and operators under the dynamical symmetry group $SU(6)\times O(3)$ \cite{Goity,PY}.  A contracted
$SU_C(6)$ spin-flavor symmetry is an emergent symmetry in the
large $N_c$ limit in baryons \cite{GS,DM},  which serves  to
organize the $1/N_c$ expansion using effective operators. The
$O(3)$ symmetry is only approximate even in large $N_c$ in the
case of the {\bf 70}-plet baryons \cite{Goity} (it becomes exact
in the case of {\bf 56}-plets \cite{Goity}), and in the large
$N_c$ limit the emergent $SU_C(6)$ is a subgroup of the
$SU(6)\times O(3)$. In the real world the breaking of
$SU(6)\times O(3)$ seems to be small, even when it happens at
${\cal{O}}(N_c^0)$, as in the case of the masses of the negative
parity baryon ${\bf 70}$-plet considered here. It is therefore
natural to implement the $1/N_c$ expansion in the framework of an
approximate $SU(6)\times O(3)$ \cite{Goity, PY, CCGL, SGS}.

Masses and partial decay widths are the main quantities
characterizing baryon resonances. These quantities can be defined and obtained through partial wave analyses where the constraints of unitarity and analyticity of the S-matrix are fulfilled. In principle they can be given 
unambiguous meaning, through  pole positions in the complex energy plane. A rigorous approach in which the $1/N_c$ expansion is implemented alongside with those analyses is yet to be developed, but it has been initiated \cite{CL}.
The analysis presented here aims at providing a  $1/N_c$  expansion for the Breit-Wigner partial decay widths, and is therefore limited in its rigor in the sense just mentioned.

The present  work extends the analysis   of  partial decay widths
of the negative parity baryons to include the decays of the
strange members of the {\bf 70}-plet as well as the decays of the
non-strange members into hyperons.  The original analysis with the
$1/N_c$ expansion was carried out in \cite{CGKM}, although an
incomplete basis of operators was used.  The improvement in the
present work, which is carried out to include $1/N_c$ corrections
and $SU(3)$ breaking to first order, is in including up to 2-body
$SU(3)$  preserving   and 1-body $SU(3)$ breaking operators. An
analysis with a complete basis at the mentioned order  is not
possible  because of the incompleteness in the input partial
widths. However, it is expected that in particular 3-body
operators are going to be dynamically suppressed.  A motivation
for the present study is to extend to the strangeness sector the
work in the non-strange sector    in the $SU(4)\times O(3)$
analysis \cite{GSS70decays}, and in particular to  determine the
importance of $SU(3)$ breaking effects.  Despite the limitations
due to the mentioned scarcity of information on strangeness
partial widths, it is possible to conclude that $SU(3)$ breaking
in partial decay widths is larger than natural size. We will be
able to show this  through the   failure of some leading
order coefficient independent relations as well as at next to
leading order where some   $SU(3)$ breaking operators are shown to
have contributions of  unnaturally large size. Particular channels
had been identified long ago \cite{FKR}, such as the S-wave
$\Lambda(1670)\to \overline{K} N$, the D-wave
$\Lambda(1690)\to \overline{K} N$, which turn out to be a problem in the
different versions of the model. The channels into $\eta$ meson
were shown to be poorly described at leading order
\cite{GSS70decays} as well.  These channels are a  most definite
proof of the large $SU(3)$ breaking effects in decays.  On the
other hand, we will show that the $SU(3)$ preserving components of
the amplitudes are well described in the $1/N_c$ expansion.

This work is organized as follows: section II presents the
framework, section III gives the construction of the bases of
operators for the partial wave decay amplitudes, section IV
presents leading order coefficient independent relations which serve
as tests of the lowest order approximation, section V presents the
fits to the Breit-Wigner partial decay widths as given in the
Particle Data Group (PDG) \cite{PDG}, and section VI is devoted to
conclusions. An appendix presents  various results needed for the
calculations of matrix elements.

\section{  1/N$_c$ expansion framework for decays}

In this section, we present the framework for implementing the
$1/N_c$ expansion for  partial decay widths of
the {\bf 70}-plet baryons, briefly reviewing  the bases 
of states and the formalism for  the decays.  Some additional  details can be found in Refs. \cite{thesis,CL2}.

The lowest lying negative parity baryons are assumed to belong to the $[{\rm
MS},\;\ell^P = 1^-]$ multiplet of $SU(6) \times O(3)$, where MS is
the $SU(6)$ mixed symmetric representation $(p,q) = (N_c-1,1)$. For $N_c=3$ this is the $[{\bf 70},1^-]$ multiplet.

 The 35 generators of $SU(6)$ along with their commutation relations are
\bea
\left[S_i,S_j\right] = i\,\epsilon_{ijk} S_k\; , &&
\left[T_a,T_b\right]  = i\, f_{abc} T_c \; , \nonumber \\
\left[S_i,G_{ja}\right]  =  i\,\epsilon_{ijk}G_{ka}\; , &&
\left[T_a,G_{ib}\right]  =  i\,f_{abc}G_{ic}\; ,\nonumber \\
\left[G_{ia},G_{jb}\right] =  i\, \delta_{ij} f_{abc} T_c & \! +\! &
i\, \epsilon_{ijk}(\delta_{ab} S_k+d_{abc} G_{ck})\; ,
\eea
where $S_i$ are the spin generators, $T_a$ the flavor generators, and $G_{ia}$ are the spin-flavor generators,
and  $d_{abc}$ and $f_{abc}$ are  respectively the  $SU(3)$ symmetric and antisymmetric  invariant tensors.

Throughout the approximation of neglecting configuration mixings
(i.e., mixing of $[{\rm MS},\; 1^-]$ with other  multiplets) will be made. This approximation is expected to
be good \cite{Mixings}, and it has to be made due to the lack of
completeness and sufficient accuracy of the  data on masses and
partial decay widths, which are  inputs to the $1/N_c$
analyses. The states in the $[{\rm MS},1^-]$ multiplet are constructed
as follows: a fundamental $SU(6)$ representation (\lq\lq excited quark\rq\rq)
is coupled to a totally symmetric (S)  representation  (\lq\lq core\rq\rq) with
baryon number $N_c-1$. The core carries spin $S_c=S+\eta$ where
$S$ is the spin of the MS $SU(6)$ state and $\eta=\pm 1/2$. The
$SU(3)$ representation of the core is determined by $S_c$  and is given by
$R_c=(p_c,q_c)=\left(2S_c, \frac{N_c-1-2S_c}{2}\right)$. The
$[{\rm MS},\;\ell]$ states then read \cite{CCGL,SGS},
\bea
\!\!\!\!\!
|J J_3;R\; (Y, I\,I_3); S >\!\!\!_{_{\rm \bf MS}} =
\sum_{\eta = \pm 1/2} C_{MS} (R,S,\eta) \
 \langle   \ell m,\, S \,  S_3  \mid   J  J_3\rangle \;
\langle   S_c\, S_{c3} ,\, \frac12 \; s_3\mid   S  S_3\rangle \; \times
\nonumber\\
\left\langle
\begin{array}{cc}
     R_c   & {\bf 3}        \\
     Y_c ~I_c ~I_{c3}     & y ~i ~ i_3
\end{array}
\right|   \left.
\begin{array}{c}
      R         \\
     Y ~I ~I_3
\end{array} \right\rangle
%\nonumber\\
|S_c\, S_{c3} ;  R_c (Y_c, I_c\, I_{c3}) > \; |\frac12 \, s_3;  {\bf 3} \; (y, i\, i_3) > \; | \ell\, m >\, ,
\eea
where   summation over repeated projection indices is implied, and $S_c=S+\eta$ in the sum.    $\ell$ is the $O(3)$ quantum number of the baryon, equal to 1 here, and
$J$ is the baryon spin. $R$ indicates  the $SU(3)$ representation of the baryon. The coefficients $C_{MS} (R,S,\eta)$  are given by \cite{SGS}:
\begin{equation}
C_{MS}(R,S,\pm \frac12) \; = \; \left\{ {\begin{array}{*{20}c}
 1 & & {\sf if} & I_R = S \pm 1 \\
0 & & {\sf if} & I_R = S \mp 1 \\
\pm \sqrt{\frac{(2S+1 \mp 1)(N_c+1\pm(2S+1))}{2N_c(2S+1)}} & & {\sf if} & I_R = S~~~~~~~~~~~~,
\end{array}} \right.
\end{equation}
where $I_R$ denotes the isospin of the  zero strangeness  states in the irreducible representation $R$ of  $SU(3)$.  For $R=(p,q)$,  $I_R=p/2$.
Later on we will indicate some of the quantum numbers of the excited baryons by an upper label ${^*}$.

For $N_c=3$  one obtains the {\bf 70}-plet $\ell =1$ states, which
consist of the following $^{2S+1}R_J$ multiplets:  $^{2}{\bf
8}_{\frac12}, ~^{2}{\bf 8}_{\frac32}, ~^{4}{\bf 8}_{\frac12},
~^{4}{\bf 8}_{\frac32},  ~^{4}{\bf 8}_{\frac52}, ~^{2}{\bf
10}_{\frac12}, ~^{2}{\bf 10}_{\frac32}, ~^{2}{\bf 1}_{\frac12},
~^{2}{\bf 1}_{\frac32}$. For generic $N_c$, the identification of
states is given in Table XV  in  the Appendix. The {\bf 1} and
{\bf 10} states have core states which for $N_c=3$ are  pure $S_c =0$
and 1 respectively. The {\bf 8}'s with same $J$ but different $S$ mix to give the mass eigenstates.
In the limit of $SU(3)$ symmetry, two mixing angles describe these possible mixings.
Due to the breaking of $SU(3)$ symmetry by the strange quark mass, the following sets of states mix with each
other: $\{ N(^{2}{\bf 8}_J),\; N(^{4}{\bf 8}_J)\}$,   $\left\{ \Sigma(^{2}{\bf 8}_J),
~\Sigma(^{4}{\bf 8}_J), ~\Sigma(^{2}{\bf 10}_J) \right\}$ , $\left\{ \Xi(^{2}{\bf 8}_J),
~\Xi(^{4}{\bf 8}_J), ~\Xi(^{2}{\bf 10}_J) \right\}$ and
$\left\{ \Lambda(^{2}{\bf 8}_J), ~\Lambda(^{4}{\bf 8}_J), ~\Lambda(^{2}{\bf 1}_J) \right\}$. The mixings, as discussed
later, can be determined primarily by the   decays and can be further constrained by the   masses  \cite{CCGL,SGS} and photo-couplings,
as it has been done in the non-strange sector
  \cite{GdeUS}. While the two mixing angles in the SU(2) symmetry limit are ${\cal O}(1)$,
the $SU(3)$  breaking effects are ${\cal O}(m_s-m_{u,d})$. 
The dimensionless $SU(3)$  breaking expansion parameter will be denoted by  $ \epsilon$, where $\epsilon \propto (m_s-m_{u,d})$.
We estimate that in  practice its value can be taken to be $\epsilon\sim 1/3$. Thus,
the mixing angles involving the octet components should differ by  ${\cal O}(\epsilon)$ corrections, while mixing
angles involving
states in different $SU(3)$  multiplets are ${\cal O}(\epsilon)$. Unfortunately, the available data on
strong decays are not sufficient to perform an analysis that can account for all the  different
mixing angles. Thus, in what follows mixing angles involving states in different $SU(3)$  multiplets
are set to vanish. The $1/N_c$ analysis of the  ${\bf 70}$-plet masses  showed  that
this is a reasonable approximation\cite{SGS}.
In this case, the states $\Sigma''_J=\Sigma(^{2}10_J)$, $\Xi''_J=\Xi(^{2}10_J)$ and
$\Lambda''_J=\Lambda(^{2}1_J)$ are taken as
unmixed states while the other physical states are obtained  as mixture between
octet states
according to
\begin{equation}
\left( \begin{array}{c}
      B_J         \\
     B'_J
\end{array} \right) = \left( \begin{array}{cc}
      \cos \theta_{B_{2J}} & \sin \theta_{B_{2J}}         \\
     -\sin \theta_{B_{2J}} & \cos \theta_{B_{2J}}
\end{array} \right) \; \left( \begin{array}{c}
      ^{2}B_J         \\
     ^{4}B_J
\end{array} \right)\, ,
\end{equation}
where for each value of $J=1/2,3/2$ we have four different angles, i.e.
$\theta_{N_{2J}}$, $\theta_{\Lambda_{2J}}$, $\theta_{\Sigma_{2J}}$ and $\theta_{\Xi_{2J}}$.

The established {\bf 70}-plet states according to the Particle
Data Group (PDG) \cite{PDG} along with their partial decay widths are
displayed in Tables I and II.  Of the known states, only the
$J=5/2$ state $N(1675)$ could have a G-wave decay into for instance
$\pi \Delta $, but no empirical information for such decay exists.
Therefore only S- and D- waves need to be considered.

The partial wave decay amplitudes via a single pseudoscalar meson can be expressed in the most general form:
\begin{equation}
 {\cal M} (\ell_{P} ~Y_{P} ~I_{P}, B_{GS}, B^*) = (-1)^{\ell_{P}} \; \sqrt{2 M_{B^*}} \;
\langle   B_{GS} \mid {\cal B}^{[\ell_{P}, R_{P}]}_{Y_{P} I_{P}} \mid B^* \rangle,
\end{equation}
where $P$ denotes a meson in the pseudoscalar octet ($R_{P}$ ={\bf
8}). $B^*$ and $B_{GS}$ are respectively the excited and ground
state baryons, $\ell_{P}$ is the partial wave, and $Y_{P}, ~I_{P}$ are
the quantum numbers of the pseudoscalar meson. A factor  involving the meson decay constant, $
 \sqrt{N_c}/F_{P}={\cal
O}(N_c^0)$, which naturally appears in the expressions
of the decay amplitude \cite{Mixings} is absorbed into the baryon operator
${\cal B}^{[\ell_{P} R_P]}_{Y_{P} I_{P}}$. This operator represents
the effective vertex $B^* B_{GS} ~P$, and will be expanded in powers of
$1/N_c$ and in $SU(3)$  breaking to first order in
$\epsilon$. The  expansion is performed with a basis of
effective operators and has the general form:
\begin{equation}
 {\cal B}^{[\ell_{P},R_{P}]}=
\left(\frac{k_{P}}{\Lambda}\right)^{\ell_{P}}
\sum_n C_n^{[\ell_{P}, R_{P}]}(k_P)\ \ {\cal B}_n^{[\ell_{P}, R_{P}]},
\end{equation}
where $k_P$ is the meson momentum, and for convenience a centrifugal barrier is factored out to
take into account the chief momentum dependence of the corresponding partial wave amplitude. The ${\cal B}_n$
represent operators in a basis, and are ordered in
powers of $1/N_c$ and $\epsilon$.  $C_n (k_P)$ are effective coefficients, which encode the QCD dynamics, and
will  be determined by fitting to the known
partial decay widths. The operators are defined and normalized such that the $C_n$'s are all  
${\cal O}(N_c^0 )$. We arbitrarily choose
the scale $\Lambda$ = 200 MeV.

\begin{table}[!h]
\caption{Empirical data for the  decay channels of the $N$'s and $\Lambda$'s
in the ${[{\bf 70},1^-]}$ from the PDG.}
\begin{center}
\begin{tabular}{ccccll}
\hline
\hline
PDG     & \hspace{.2cm} State \hspace{.2cm} &  Mass   & Total Width& \multicolumn{2}{c}{Branching ratios [\%]}\\
Name    &                                   &   [MeV] & [MeV]  &
\hspace{.5cm}S-wave\hspace{.5cm} &
\hspace{.5cm}D-wave\hspace{.5cm}    \\
\hline
\hline
$N(1535)$ & $N_{1/2}$    & $1535(10)$ & $150(25)$     & $\pi N$  :  $45(10)$        &  $\pi \Delta$ $<1$   \\
          &               &        &              & $\eta N$ : $52.5(7.5)$ &                      \\[0.5mm]
\hline
$N(1520)$ & $N_{3/2}$    & $1520(5)$ & $113(12.5)$ & $\pi \Delta$ :  $8.5(3.5)$ & $\pi N$ : $60(5)$       \\
          &              &           &             &                          & $\pi \Delta$ : $12(2)$  \\
          &              &           &             &                          & $\eta N$ : $0.23(0.04)$  \\[0.5mm]
\hline
$N(1650)$ &  $N'_{1/2}$  & $1657(13)$ & $165(20)$ &  $\pi N$      :  $77.5(17.5)$ & $\pi \Delta $ : $4(3)$   \\[0.5mm]
          &              &        &               &  $\eta N$     : $6.5(3.5)$   &                         \\
          &              &        &               &  $K \Lambda$  : $7(4)$       &                         \\[0.5mm]
\hline
$N(1700)$ &  $N'_{3/2}$   & $1700(50)$ & $100(50)$    &          $\pi \Delta$ :  $90(5)$                 &   $\pi N$     : $10(5)$
\\
        
          &               &        &              &                        & $K \Lambda$ $<3$     \\[0.5mm]
\hline
$N(1675)$ &  $N_{5/2}$   & $1675(5)$ & $148(18)$ &                        & $\pi N$      : $40(5)$  \\
          &              &        &               &                        & $K \Lambda$   $<1$      \\[0.5mm]
\hline
$\Lambda(1670)$ & $\Lambda_{1/2}$   & $1670(10)$ & $37.5(12.5)$ & $\bar K N$     : $25(5)$     &   \\
                &                   &        &              & $\eta \Lambda$ : $17.5(7.5)$ &   \\
                &                   &        &              & $\pi \Sigma $   : $40(15)$    &   \\
\hline
$\Lambda(1690)$ &  $\Lambda_{3/2}$   & $1690(5)$ & $60(10)$ & $\pi \Sigma^*$ : $45(10)$  & $\bar K N$   : $25(5)$ \\
                &     &        &          &  & $\pi \Sigma$ : $30(10)$
   \\[0.5mm]
\hline
$\Lambda(1800)$ & $\Lambda'_{1/2}$  & $1785(65)$ & $300(100)$   & $ \overline{K}N $     : $32.5(7.5)$ &   \\
\hline
$\Lambda(1830)$ &  $\Lambda_{5/2}$    & $1820(10)$ & $85(25)$ &  & $\bar K N$         : $6.5(3.5)$    \\
                &     &        &          &  & $\pi \Sigma$       : $55(20)$    \\
                &     &        &          &  & $\pi \Sigma^*$      $>15$        \\[0.5mm]
\hline
$\Lambda(1405)$ & $\Lambda''_{1/2}$ & $1406(4)$ & $50(2)$      & $\pi \Sigma$   : $100$       &        \\
\hline
$\Lambda(1520)$ &  $\Lambda''_{3/2}$    & $1519(1)$ & $15.6(1)$ & &  $\bar K N$        : $45(1)$  \\
                &     &        &           & &  $\pi \Sigma$      : $42(1)$    \\[0.5mm]
\hline
\end{tabular}
\end{center}
\end{table}
\begin{table}[!h]
\caption{Empirical data for decay channels of the $\Sigma$'s and $\Delta$'s in the ${[{\bf 70},1^-]}$ from the PDG.}
\begin{center}
\begin{tabular}{cccccccc}
\hline
\hline
PDG     & \hspace{.2cm} State \hspace{.2cm} &  Mass   & Total Width& \multicolumn{2}{c}{Branching ratios}\\
Name    &                                   &   [MeV] & [MeV]  &
\hspace{.5cm}S-wave\hspace{.5cm} &
\hspace{.5cm}D-wave\hspace{.5cm}    \\
\hline
\hline
$\Sigma(1670)$  &  $\Sigma_{3/2}$    & $1675(10)$ & $60(20)$  &  & $\bar K N$        : $10(3)$    \\
                &                    &        &           &  & $\pi \Lambda$     : $10(5)$     \\
                &                    &        &           &  & $\pi \Sigma$      : $45(15)$   \\[0.5mm]
\hline
$\Sigma(1750)$  & $\Sigma'_{1/2}$   & $1765(35)$ & $110(50)$    & $\bar K N$     : $25(15)$   &    \\
                &                   &        &              & $\pi \Sigma$    $<8$    &        \\
                &                   &        &              &  $\eta \Sigma$  : $35(20)$  &   \\[1.5mm]
\hline
$\Sigma(1775)$  &  $\Sigma_{5/2}$   & $1775(5)$ & $120(15)$    &    & $\bar K N$        : $40(3)$    \\
                &                   &        &              &    & $\pi \Lambda$     : $17(3)$      \\
                &                   &        &              &    &  $\pi \Sigma$     : $3.5(1.5)$  \\
                &                   &        &              &    & $\pi \Sigma^*$    : $10(2)$     \\[0.5mm]
\hline
$\Delta(1620)$  &  $\Delta_{1/2}$   & $1630(30)$ & $143(7.5)$ &  $\pi N$  : $25(5)$ & $\pi \Delta$ : $45(15)$ \\[0.5mm]
\hline
$\Delta(1700)$  &  $\Delta_{3/2}$    & $1710(40)$ & $300(100)$ & $\pi \Delta$   : $37.5(12.5)$ & $\pi N$ : $15(5)$  \\
                &     &        &            &    & $\pi \Delta$ : $4(3)$  \\[0.5mm]
\hline \hline
\end{tabular}
\end{center}
\end{table}

The basis operators are expressed in terms of spin-flavor operators ${\cal{G}}_n$:
\beq
{\cal B}^{[\ell_{P}, R_{P}]}_n = \left( \xi^{\ell} \; {\cal G}^{[j_n, R_{P}]}_n \right)^{[\ell_{P}, R_{P}]},
\eeq
with the obvious notation indicating coupling of angular momenta.  $\xi^{\ell}$ is an $O(3)$
tensor operator for the transition from
the $O(3)$  state with $\ell=1$ of  the excited baryon to the GS baryon where $\ell$ = 0.   ${\cal G}_n$ is a spin-flavor operator,
which gives the transition from the initial $SU(6)$  MS baryon state to the GS baryon which is a symmetric state.
Without any loss of generality, one can choose
$\xi^{\ell}$ to satisfy $< 0 |\xi^{\ell}_{m'} | \ell ~m > = (-1)^{\ell - m} \; \delta_{m,-m'}$.

The  partial decay width is then given in terms of the reduced matrix elements (RMEs) of the operators ${\cal B}_n$
as follows:
\bea \label{width}
\Gamma_{\ell_{P} Q_{P}}&=& \frac{k_{P}}{8\pi^2}
\left(\frac{k_{P}}{\Lambda}\right)^{2 \ell_{P}}\frac{M_B}{M^*_B}
\;\frac{\hat{I}^2}{(\hat{I^*} \hat{J^*})^2} \\
& &\!\!\!\!\!\!\!\!\!\!\!\!\!\!\!\!\!\! \times
\Big| \sum_{n} C_n^{[\ell_{P}, R_{P}]}(k_P)
\sum_{\gamma}
\left(\begin{array}{cc}
     R^*   & R_{P}        \\
     Y^*~I^*     & Y_{P} ~I_{P}
\end{array} \right\Arrowvert   \left.  \begin{array}{c}
      R         \\
     Y~~ I
\end{array} \right)_{\gamma}
 \;{\cal B}_n^{\ \gamma}(\{S,Q\},\{(\ell,S^*)J^*,Q^*\},\{\ell_{P}, Q_{P}\} )    \Big|^2,\nonumber
\eea
where throughout we use the customary notation for the $SU(3)$  isoscalar factors,
$\hat{t}\equiv \sqrt{2\ t+1}$,  and
$Q\equiv\{R,Y,I\}$. In addition, $\gamma$ labels the possible multiplicities
in the coupling $R^* \otimes R_{P} \to R$. Since there is mixing between $S^*$ = 1/2 and 3/2 states in
the case $R^*$ = {\bf 8},
this mixing is simply taken into account by replacing ${\cal B}_n^{\ \gamma}$ in the
formula by the corresponding linear combination with the $S^*$ = 1/2 and $S^*$ = 3/2 states, and similarly
for the $SU(3)$ breaking induced
mixings involving states in different $R^*$ representations, i.e., {\bf 1},  {\bf 8} and {\bf 10} if one would include those mixings.

Distinguishing the operators into $SU(3)$  preserving and $SU(3)$  breaking, the RMEs
can be expressed in terms of the RMEs of the  spin-flavor operators ${\cal G}_n$. For $SU(3)$ preserving operators
one has:
\bea
& &{\cal B}_n^{\ \gamma} (\{S,Q\},\{(\ell,S^*)J^*,Q^*\},
\{\ell_{P},Q_{P}\})=\nonumber\\
&& (-1)^{j_n+J^*+\ell+S}\;\frac{\hat{J^*}\;\hat{\ell_{P}}}{\sqrt{{\rm dim}\; R}}
\left\{
 \begin{array}{ccc}
     J^*   & S^*     & \ell   \\
     j_n    & \ell_{P}     & S
\end{array} \right\} \;
\langle S,R\| {\cal{G}}_n^{[j_n,R_{P}]}\|S^*,R^*\rangle_\gamma~~,
\eea
where one identifies a 6-j $SU(2)$  symbol and the  RME of the spin-flavor operator.
We use the $SU(2)$ conventions from Edmonds \cite{Edmonds} and the $SU(3)$ conventions
from Hecht \cite{Hecht}.

To first order
in the quark masses, the $SU(3)$  symmetry breaking can be expressed in terms of
spin-flavor operators  of the form:
\beq
 \left({\cal{M}}^8_f \; {\cal{G}}_m^{[j_m, R_{m}]}\right)_{\gamma_n}^{R_{P}},
 \eeq
where ${\cal{M}}^8_f$ is the octet component of the quark masses (we work in the limit of exact isospin symmetry). Since
in the present case $R_{P} ={\bf  8}$, $R_m$ can be {\bf 1}, {\bf 8}, {\bf 10},  ${\bf \overline {10}}$ or {\bf 27}. Only $R_m = {\bf 1},~{\bf 8}$ involve 1-body spin-flavor operators, while
 {\bf 10},  ${\bf \overline {10}}$ and  {\bf 27} involve 2-body operators. There is, therefore, in principle a significant number of $SU(3)$ breaking operators. However, the reality of the matter
is  that the available information on partial decay widths sensitive to  $SU(3)$  breaking effects is very limited, and for this reason, in the present work,
only a truncated basis of such operators will be used, namely 1-body ones.

Up to an overall constant to be absorbed by the normalization procedure discussed below,
the RMEs of  $SU(3)$  breaking operators can be shown to be of the
most general form \cite{JGS}
\bea
{{\cal B}}_n^{\ \gamma}(\{S,Q\},\{(\ell,S^*)J^*,Q^*\},
\{\ell_{P},Q_{P}\}) &=&
(-1)^{j_m+J^*+\ell+S}\;\frac{\hat{J^*}\;\hat{\ell_{P}}}{\sqrt{{\rm
dim}\; R}} \left\{
\begin{array}{ccc}
     J^*   & S^*     & \ell   \\
     j_m     & \ell_{P}     & S
\end{array} \right\} \ \times
\nonumber \\
& &  \left( \begin{array}{cc}
     8   &  R_m        \\
     0~0     &  Y_{P}~ I_{P}
\end{array}  \right\Arrowvert \left. \begin{array}{c}
      R_{P}         \\
   Y_P~ I_{P}
\end{array}\right)_{\!\!\gamma_n}\!\!\!
\langle S,R\| {\cal{G}}_m^{[j_m, R_m]}\|S^*,R^*\rangle_{\gamma}~.
\nonumber \\
\eea

Basis operators will be normalized such that the coefficients $C_n$ are all
${\cal O}(N_c^0\times \epsilon^0)$ and their  natural size within a given partial wave be the same. In this manner,
just by looking at the size of a coefficient, one can infer whether the corresponding operator is giving contributions within the expectations of its
$1/N_c$ and $\epsilon$ power countings.   Effects of $SU(6)$ symmetry breaking are reflected in the momentum $k_P$, which   are then taken into account    by the decay operators and are  therefore   encoded in the effective coefficients  $C_n$. In the case of the ${\bf 56}$-plet decays, where all $SU(6)$ breaking effects are   ${\cal O}(1/N_c)$ or  ${\cal O}(\epsilon)$, the $k_P$ dependencies are sub-leading in the respective expansions.
For this reason such effects are simply taken into account by the expansion itself if one neglects the $k_P$ dependence of the coefficients $C_n$.  In the  case of the
  ${\bf 70}$-plet, the
${\cal O}(N_c^0)$  effects on  mass splittings are small, and thus a similar approach of neglecting the momentum dependency of the coefficients is expected to work.

\section{  Operator Bases}

The construction of the bases of operators  was already discussed in \cite{GSS70decays} for the case of $SU(4)$, and is briefly reviewed here. The
spin-flavor transition operators ${\cal{G}}_n$  from  the  MS to S  representations    are built using tensor products of the $SU(6)$ generators $\Lambda_c$ and $\lambda$, which operate on the \lq\lq core\rq\rq and
\lq\lq excited quark\rq\rq  respectively. The reduction rules established in \cite{DJM} are used to reduce products of core generators, and products ($\alpha, ~\beta,\cdots$,  $SU(6)$ generator indices) $\lambda_\alpha \lambda_\beta $ can be reduced to 1-body operators. In addition, using   that any generator of $SU(6)$  $\Lambda = \Lambda_c + \lambda$ has vanishing matrix elements between MS and  S  states,
one obtains the additional equivalences:

\bea
&1-{\rm body}& \hspace{1cm} \Lambda_{c\; \alpha}= -\lambda_{\alpha} \\
&2-{\rm body}& \hspace{1cm} \Lambda_{c\;\alpha} \; \Lambda_{c \;\beta} = -\lambda _\alpha ~\Lambda_{c\;\beta} - \lambda _\beta ~\Lambda_{c\;\alpha} + {\rm 1}\!-\!{\rm body~operators}. \nonumber
\eea
Therefore, only the following types of operators need to be considered,
\bea
&1-{\rm body}& \hspace{1cm} \lambda \nonumber\\
&2-{\rm body}& \hspace{1cm} \frac{1}{N_c} ~\lambda_\alpha ~\Lambda_{c\;\beta} \nonumber\\
&3-{\rm body}& \hspace{1cm} \frac{1}{N_c^2} ~\lambda_\alpha ~\Lambda_{c\;\beta} ~\Lambda_{c\;\gamma} \; .
\eea

There is only one 1-body operator, namely
\begin{eqnarray}
{\cal G}^{(1)} = g^{[1,8]},
\end{eqnarray}
which is identified with the axial current of the "excited quark" in quark model language. On the other hand 2-body operators can be
built with products $\lambda \, \Lambda_c$,  which can be coupled to angular momentum  $j=0,~1,~2$. For  $\ell=1$,  $\ell_{P}$ can be $0,~2,~4$,  therefore  $j=1,~2$. They have to be coupled to an {\bf 8} of $SU(3)$  for the decays involving  pseudoscalar mesons in $R_{P}={\bf 8}$.  The following possible monomials can appear in the 2-body operators:
\begin{equation}
\begin{array}{l}
\left( s~T^c \right)^{[1,8]},\; \left( t~S^c \right)^{[1,8]},\; \left( s~G^c \right)^{[j,8]}, \; \left( g~S^c \right)^{[j,8]}  ,\\
\left( t~G^c \right)^{[1,8]}, \; \left( g~T^c \right)^{[1,8]},\; \left( g~G^c \right)^{[j,8]},  
\end{array}
\end{equation}
where $j=1,2$, and the coupling of the two $\bf  8$s  to  $\bf  8$  involves two
possibilities, namely  $f$ and $d$ type couplings.  There are four such cases in this  list of operators, and therefore  the total number of possible  monomials  is equal to fourteen.

This set of 2-body operators can be reduced using   operator identities, resulting in nine independent operators, namely:
\begin{equation}
\begin{tabular}{l l l l l}
${\cal G}_1^{(2)} = \frac{1}{N_c} \left( s T^c \right)^{[1,8]}$
&\hspace*{0.35cm} &
${\cal G}_2^{(2)} = \frac{1}{N_c} \left( t S^c \right)^{[1,8]}$
&\hspace*{0.35cm} &
${\cal G}_3^{(2)} = \frac{1}{N_c} \left( s G^c \right)^{[1,8]} $
\\[0.25cm]
${\cal G}_4^{(2)} = \frac{1}{N_c} \left( g  S^c\right)^{[1,8]}$
& &
${\cal G}_5^{(2)} = \frac{1}{N_c} \left( g T^c  \right)^{[1,8]}_{\gamma=1}$
& &
${\cal G}_6^{(2)} = \frac{1}{N_c} \left( g T^c  \right)^{[1,8]}_{\gamma=2}$
\\[0.25cm]
${\cal G}_7^{(2)} = \frac{1}{N_c} \left( s G^c \right)^{[2,8]}$
& &
${\cal G}_8^{(2)} = \frac{1}{N_c} \left( g S^c\right)^{[2,8]}$
& &
${\cal G}_9^{(2)} = \frac{1}{N_c} \left( g G^c\right)^{[2,8]}_{\gamma=1}$ .
\end{tabular}
\end{equation}
So far, we have discussed the operators that appear in the limit of exact $SU(3)$ symmetry.

As  mentioned earlier in the  $SU(3)$  breaking operators,  Eqn.(\ref{breakers}), ${\cal{G}}_m$ belong to
{\bf 1}, {\bf 8}, {\bf 10}, ${\bf \overline{10}}$ or {\bf 27}  of
$SU(3)$.     The operators of interest here are  the 1-body ones, which are ${\cal O}(\epsilon
\times N_c^0)$. The basis of  1-body  $SU(3)$  breaking operators is constructed from the following three monomials:
\bea
{\cal G}_1^{(SB)} &=& \left( t_8 s \right)^{[1,8]} \nonumber \\
{\cal G}_2^{(SB)} &=& \left( t_8 g \right)^{[1,8]}_{\gamma=2} - \sqrt{\frac{2}{3}} \ {\cal G}^{(1)} \nonumber\\
{\cal G}_3^{(SB)} &=& \left( t_8 g \right)^{[1,8]}_{\gamma=1}.
\label{breakers}
\eea
Note that ${\cal G}_1^{(SB)} $ only couples to the $\eta$ meson.
For convenience ${\cal G}_2^{(SB)}$ has been defined in such a way
that channels with emission of a $\pi$ meson are not affected.
${\cal G}_3^{(SB)}$ automatically has that property. For a list of
the 2-body symmetry breaking operators, see \cite{thesis}. 

Three
body basis operators can be constructed following similar procedure. One
would however expect that they will be dynamically suppressed, as it has been noted in the particular case of mass operators.

The list of basis operators used in this work is shown in Table
III.
 Throughout  it is  assumed   that the number of strange quarks in
the baryons is ${\cal O}(N_c^0)$, while of course the number of
$u$ and $d$ quarks is assumed to be ${\cal O}(N_c)$. This reflects
in the counting of powers of $1/N_c$ in the decay amplitudes
involving strange baryons. Since the only strange baryons included as inputs in the analysis    have strangeness (-1), it seems that the
assumption is reasonable. We  have then the following general
$1/N_c$ power countings: the amplitudes of non-strange baryons
decaying via $\pi$ or $\eta$ meson emission are all ${\cal
O}(N_c^0)$ up to corrections ${\cal O}(1/N_c)$, while their decay
amplitudes via $K$-meson emission are ${\cal O}(1/\sqrt{N_c})$.
Similarly, the decay amplitudes of strange baryons via $\pi$
emission are ${\cal O}(N_c^0)$, and via $K$-meson  emission are ${\cal
O}(1/\sqrt{N_c})$ except for the singlet $\Lambda(1405)$ and
$\Lambda(1520)$ baryons where it is reversed. These properties can
be traced back to the $1/N_c$ dependence of isoscalar factors. A
similar situation occurs with the $SU(3)$ breaking operators used
here, except that for these    the $\pi$-emission amplitudes are
identically zero.
\begin{table}[!h]
\caption{The effective operators for ${\bf 70}$-plet baryon decays. First operator set represents the symmetric operators and second set represents
the $SU(3)$ symmetry breaking operators. $\ell=0$ for S-wave and $\ell=2$ for D-wave decays.}
\begin{center}
\begin{tabular}{l l c c c c}
\hline \hline
& \hspace{2cm}Operator & & n-bodyness & & Order \\
\hline \hline
&${\cal B}_1 \; = \; \left(\xi \ \! g\right)^{[\ell,8]}$ & & 1 & &  $N_c^0$  \\[1mm]
&${\cal B}_2 \; = \; \frac{1}{N_c} \left(\xi \left(s \ \! T^c \right)^{[1,8]} \right)^{[\ell,8]}$ & & 2 & & $N_c^0$ \\[1mm]
&${\cal B}_3 \; = \; \frac{1}{N_c} \left(\xi \left(t \ \! S^c \right)^{[1,8]} \right)^{[\ell,8]}$ & & 2 & & $1/N_c$ \\[1mm]
&${\cal B}_4 \; = \; \frac{1}{N_c} \left(\xi \left(s \ \! G^c \right)^{[1,8]} \right)^{[\ell,8]}$ & & 2 & &  $N_c^0$  \\[1mm]
&${\cal B}_5 \; = \; \frac{1}{N_c} \left(\xi \left(g \ \! S^c \right)^{[1,8]} \right)^{[\ell,8]}$ & & 2 & & $1/N_c$ \\[1mm]
&${\cal B}_6 \; = \; \frac{1}{N_c} \left(\xi  \left(g \ \! T^c  \right)^{[1,8]}_{\gamma=1} \right)^{[\ell,8]}$ & & 2 & &  $N_c^0$  \\[1mm]
&${\cal B}_7 \; = \; \frac{1}{N_c} \left(\xi  \left(g \ \! T^c  \right)^{[1,8]}_{\gamma=2} \right)^{[\ell,8]}$ & & 2 & &  $N_c^0$  \\[1mm]
&${\cal B}_8 \; = \; \frac{1}{N_c} \left(\xi \left(s \ \! G^c \right)^{[2,8]} \right)^{[\ell,8]}$ & & 2 & &  $N_c^0$  \\[1mm]
&${\cal B}_9 \; = \; \frac{1}{N_c} \left(\xi \left(g \ \! S^c \right)^{[2,8]} \right)^{[\ell,8]}$ & & 2 & & $1/N_c$ \\[1mm]
&${\cal B}_{10} \; = \; \frac{1}{N_c} \left(\xi \left(g \ \! G^c \right)^{[2,8]}_{\gamma=1} \right)^{[\ell,8]}$ & & 2 & &  $N_c^0$  \\[2mm]
\hline \hline \\[-0.5cm]
&${\cal B}_1^{SB} \; = \; \left(\xi \left( t_8 \ \! s \right)^{[1,8]}_{\gamma=1}\right)^{[\ell,8]}$ & & 1 & & $\epsilon$  \\[1mm]
&${\cal B}_2^{SB} \; = \; \left(\xi \left( t_8 \ \! g \right)^{[1,8]}_{\gamma=2}\right)^{[\ell,8]} - \sqrt{\frac{2}{3}}\ {\cal B}_1$& & 1 & & $\epsilon$ \\[1mm]
&${\cal B}_3^{SB} \; = \; \left(\xi \left( t_8 \ \! g \right)^{[1,8]}_{\gamma=1}\right)^{[\ell,8]}$ & & 1 & & $\epsilon$  \\[2mm]
\hline \hline
\end{tabular}
\end{center}
\end{table}

For arbitrary $N_c$  the number of $SU(3)$  preserving  basis operators  is equal
to   seven  for S-waves and  to ten  for D-waves. The   RMEs  ${\bf {\cal B}}_n^{\
\gamma}$ are  shown in Tables IV and V. The tables  also show  the normalization
factors ($\alpha_{n}$)   necessary to  have each operator give natural size
matrix elements. As explained earlier, these are chosen in such a way that the
largest spin-isospin RME of ${\cal O}(N_c^0)$ operators is equal to one in magnitude when
evaluated at $N_c$ = 3, and 1/3 for the case of  ${\cal O}(1/N_c)$ $SU(3)$
preserving operators. Similarly, the  latter  normalization is implemented in
$SU(3)$  breaking operators as well. 
Note that the spin-isospin RMEs are defined by the sum
over $\gamma$ of the product of $B_n^\gamma$ with the corresponding isoscalar factor
(see Eqn.(\ref{width})).

%\newpage

\begin{table}[!h]
\caption{Reduced matrix elements ${\bf {\cal B}}_n^{\ \gamma}$ (see Eqn. (8)) for S-wave decays}
\begin{center}
\small{
\begin{tabular}{c|ccccccccccccccr}
\hline \hline
Decay & ${\bf {\cal B}}_1$&& ${\bf {\cal B}}_2$&& ${\bf {\cal B}}_3$&& ${\bf {\cal B}}_4$&& ${\bf {\cal B}}_5$&& ${\bf {\cal B}}_6$&&  ${\bf {\cal B}}_7$&&  Overall\\[-0.2cm]
chanel&   &&  &&  &&  &&  &&  && &&~factor \\
\hline \hline
$(^{2}8_{\frac12} \to 8)_{_1}$ & $-\frac{12+N_c}{6\sqrt{2}}$ && $\frac{\beta_1}{2\sqrt{2}}$ && $\frac{3+N_c}{6\sqrt{2}N_c}$ && $-\frac14$ && 0 && 0 && $\frac{\beta_2}{\sqrt{8}} $  &&$A_2$\\[1.5mm]
$(^{2}8_{\frac12} \to 8)_{_2}$ & $\frac{2N_c-3}{6\sqrt{2}}$ && $\frac{3-2N_c}{6\sqrt{2}N_c}$ && $\frac{3+N_c}{6\sqrt{2}N_c}$ && $\frac{N_c}{12}$ && 0 &&0&&   $\frac{\beta_3}{6\sqrt{2}} $  &&$A_3$\\[1.5mm]
$^{2}8_{\frac32} \to 10$ & $-\frac{1}{6\sqrt{2}}$ && $\frac{1}{6 \sqrt{2}N_c}$ && $\frac{1}{6\sqrt{2}N_c}$ && $\frac{3-N_c}{24N_c}$ && $-\frac{1}{8N_c}$ &&$-\frac{1}{4\sqrt{6}N_c}$ && $\frac{N_c-2}{12\sqrt{30} N_c}$  &&$-A_1$\\[1.5mm]
$(^{4}8_{\frac12} \to 8)_{_1}$ & $\frac16$ && $\frac{2+N_c}{6N_c}$ && $-\frac{1}{6N_c}$ && $-\frac{1}{2\sqrt{2}N_c}$ && $-\frac{1}{4\sqrt{2}N_c}$ &&0&&   $\frac{N_c+17}{12 \sqrt{15} N_c}$   && $-A_5$ \\[1.5mm]
$(^{4}8_{\frac12} \to 8)_{_2}$ & $\frac16$ && $-\frac{1}{6N_c}$ && $-\frac{1}{6N_c}$ && $\frac{N_c-3}{12\sqrt{2}N_c}$ && $-\frac{1}{4\sqrt{2}N_c}$ && 0 &&     $\frac{11-N_c}{12 \sqrt{15} N_c}$    && $-A_6$ \\[1.5mm]
$^{4}8_{\frac32} \to 10$ & $\frac16\sqrt{\frac52}$ && $-\frac{1}{6N_c}\sqrt{\frac52}$ && $\frac{1}{3N_c}\sqrt{\frac52}$ && $\frac{\sqrt{5}(N_c-3)}{24N_c}$ && 0 && $\frac{1}{4N_c}\sqrt{\frac56}$ && $\frac{2-N_c}{12 \sqrt{6} N_c}$   &&$A_4$ \\[1.5mm]
$^{2}10_{\frac12} \to 8$ & $\frac16$ && $-\frac{1}{6N_c}$ && $-\frac{2}{3N_c}$ && $\frac{N_c+3}{12\sqrt{2}N_c}$ && 0 && $-\frac{1}{4\sqrt{3}N_c}$ && $\frac{2-N_c}{12 \sqrt{15} N_c}$ && $A_9$ \\[1.5mm]
$(^{2}10_{\frac32} \to 10)_{_1}$ & $\frac{21+N_c}{6\sqrt{2}}$ && $\frac{\beta_4}{2\sqrt{2}}$ && $-\frac{21+N_c}{6\sqrt{2}N_c}$ && $\frac{3(N_c+1)}{4N_c}$ && $\frac{21+N_c}{8N_c}$ && 0 && $\frac{\beta_5}{\sqrt{8}}$ && $A_7$ \\[1.5mm]
$(^{2}10_{\frac32} \to 10)_{_2}$ & $\frac16\sqrt{\frac52}$ && $-\frac{1}{6N_c}\sqrt{\frac52}$ && $-\frac{1}{6N_c}\sqrt{\frac52}$ && $\frac{\sqrt{5}(Nc+3)}{24N_c}$ && $\frac{\sqrt{5}}{8N_c}$ && 0 && $-\frac{N_c+7}{12 \sqrt{6}N_c}$ && $-A_8$ \\[1.5mm]
$^{2}1_{\frac12} \to 8$ & $-\frac14$ && $\frac{1}{4N_c}$ && 0 && $\frac{1}{4\sqrt{2}N_c}$ && 0 && $-\frac{N_c+3}{16\sqrt{3}N_c}$ &&     $-\frac{N_c+7}{16\sqrt{15}N_c}$ && $A_5$ \\[2mm]\hline
$\alpha_{n}$ & $\sqrt{\frac{27}{10}}$ && $\frac{9}{2}\sqrt{\frac 35}$ && $\sqrt{\frac{27}{40}}$ && $2\sqrt{\frac 65}$ && $6\sqrt{\frac 65}$ && $9\sqrt{\frac{2}{5}}$ && $\frac{27}{11}\sqrt{10}$ && \\[2mm]
\hline \hline
\end{tabular}}
\end{center}
\vspace{-.35cm}
\end{table}
\begin{table}[!h]
\caption{Reduced matrix elements ${\bf {\cal B}}_n^{\ \gamma}$ (see Eqn. (8)) for D-wave decays.}
\begin{center}
{\scriptsize
\begin{tabular}{c|ccccccccccr}
\hline \hline
Decay & ${\bf {\cal B}}_1$ &${\bf {\cal B}}_2$ &${\bf {\cal B}}_3$ &${\bf {\cal B}}_4$ &${\bf {\cal B}}_5$&
${\bf {\cal B}}_6$ &${\bf {\cal B}}_7$ &${\bf {\cal B}}_8$ & ${\bf {\cal B}}_9$ &${\bf {\cal B}}_{10}$ & Overall\\[-0.2cm]
Channel & & & & & & & & & &  & ~factor\\
\hline \hline
& & & & & & & & & & & \\[-1.5mm]
$^{2}8_{\frac12} \to 10$&$\frac13$ &$-\frac{1}{3N_c}$ &$-\frac{1}{3N_c}$ &$\frac{N_c-3}{6\sqrt{2}N_c}$ &$\frac{1}{2\sqrt{2}N_c}$ &
$\frac{1}{2\sqrt{3}N_c}$ &$\frac{2-N_c}{6\sqrt{15}N_c}$ &$\frac{1-N_c}{2\sqrt{6}N_c}$ &$\frac{1}{2\sqrt{6}N_c}$ &
$\frac{1}{12\sqrt{2}}$ & $\frac{\sqrt{5}}{4}A_1$\\[1.5mm]
$(^{2}8_{\frac32} \to 8)_{_1}$&$-\frac{N_c+12}{3}$ &$\beta_1$ &$\frac{N_c+3}{3N_c}$ &$-\frac{1}{\sqrt{2}}$ &0&0&$\beta_2$
 &0&0&0&$-\frac{\sqrt{10}}{4}A_2$ \\[1.5mm]
$(^{2}8_{\frac32} \to 8)_{_2}$&$2N_c-3$ &$\frac{3-2N_c}{N_c}$ &$\frac{3+N_c}{N_c}$ &$\frac{N_c}{\sqrt{2}}$ &0&0&$\beta_3$
 &0&0&0&$-\frac{\sqrt{10}}{12}A_3$ \\[1.5mm]
$^{2}8_{\frac32} \to 10$&$\frac13$ &$-\frac{1}{3N_c}$ &$-\frac{1}{3N_c}$ &$\frac{N_c-3}{6\sqrt{2}N_c}$ &$\frac{1}{2\sqrt{2}N_c}$ &
$\frac{1}{2\sqrt{3}N_c}$ &$\frac{2-N_c}{6\sqrt{15}N_c}$ &$\frac{N_c-1}{2\sqrt{6}N_c}$ &$-\frac{1}{2\sqrt{6}N_c}$ &
$-\frac{1}{12\sqrt{2}}$ & $\frac{\sqrt{5}}{4}A_1$\\[1.5mm]
$^{4}8_{\frac12} \to 10$&$\frac13$ &$-\frac{1}{3N_c}$ &$\frac{2}{3N_c}$ &$\frac{N_c-3}{6\sqrt{2}N_c}$ &0&$\frac{1}{2\sqrt{3}N_c}$&
$\frac{2-N_c}{6\sqrt{15}N_c}$ &$\frac{1-N_c}{2\sqrt{6}N_c}$ &$\sqrt{\frac23}\frac{1}{N_c}$ &
$\frac{3+N_c}{12\sqrt{2}N_c}$ & $-\frac{\sqrt{10}}{8}A_4$\\[1.5mm]
$(^{4}8_{\frac32} \to 8)_{_1}$&-1 &$-\frac{N_c+2}{N_c}$ &$\frac{1}{N_c}$ &$\frac{3}{\sqrt{2}N_c}$ &$\frac{3}{2\sqrt{2}N_c}$
 &0&$-\frac{N_c+17}{2\sqrt{15}N_c}$ &$-\frac{3\sqrt{6}}{N_c}$ &$-\sqrt{\frac32}\frac{3}{2N_c}$ &
$\frac{9}{4\sqrt{2}N_c}$ & $\frac{1}{6\sqrt{2}}A_5$ \\[1.5mm]
$(^{4}8_{\frac32} \to 8)_{_2}$&-1 &$\frac{1}{N_c}$ &$\frac{1}{N_c}$ &$\frac{3-N_c}{2\sqrt{2}N_c}$ &$\frac{3}{2\sqrt{2}N_c}$
 &0&$\frac{N_c-11}{2\sqrt{15}N_c}$ &$\sqrt{\frac32}\frac{3(N_c-1)}{2N_c}$ &$-\sqrt{\frac32}\frac{3}{2N_c}$ &
$-\frac{3}{4\sqrt{2}}$ & $\frac{1}{6\sqrt{2}}A_6$ \\[1.5mm]
$^{4}8_{\frac32} \to 10$&-1 &$\frac{1}{N_c}$ &$-\frac{2}{N_c}$ &$\frac{3-N_c}{2\sqrt{2}N_c}$ &0&$-\frac{\sqrt{3}}{2N_c}$ &$\frac{N_c-2}{2\sqrt{15}N_c}$ &
$\sqrt{\frac32}\frac{N_c-1}{4N_c}$ &$-\sqrt{\frac32}\frac{1}{N_c}$ &$-\frac{N_c+3}{8\sqrt{2}N_c}$
& $\frac13A_4$ \\[1.5mm]
$(^{4}8_{\frac52} \to 8)_{_1}$&1 &$\frac{N_c+2}{N_c}$ &$-\frac{1}{N_c}$ &$-\frac{3}{\sqrt{2}N_c}$ &$-\frac{3}{2\sqrt{2}N_c}$
 &0&$\frac{N_c+17}{2\sqrt{15}N_c}$ &$-\sqrt{\frac23}\frac{1}{N_c}$ &$-\frac{1}{2\sqrt{6}N_c}$ &$\frac{1}{4\sqrt{2}N_c}$
&$-\frac{1}{2\sqrt{2}}A_5$ \\[1.5mm]
$(^{4}8_{\frac52} \to 8)_{_2}$&1 &$-\frac{1}{N_c}$ &$-\frac{1}{N_c}$ &$\frac{N_c-3}{2\sqrt{2}N_c}$ &$-\frac{3}{2\sqrt{2}N_c}$
 &0&$\frac{11-N_c}{2\sqrt{15}N_c}$ &$\frac{N_c-1}{2\sqrt{6}N_c}$ &$-\frac{1}{2\sqrt{6}N_c}$ &$-\frac{1}{12\sqrt{2}}$ &
$-\frac{1}{2\sqrt{2}}A_6$ \\[1.5mm]
$^{4}8_{\frac52} \to 10$&1 &$-\frac{1}{N_c}$ &$\frac{2}{N_c}$ &$\frac{N_c-3}{2\sqrt{2}N_c}$ &0&$\frac{\sqrt{3}}{2N_c}$ &$\frac{2-N_c}{2\sqrt{15}N_c}$ &
$\frac{N_c-1}{2\sqrt{6}N_c}$ &$-\sqrt{\frac23}\frac{1}{N_c}$ &$-\frac{N_c+3}{12\sqrt{2}N_c}$
&$-\frac{\sqrt{14}}{8}A_4$ \\[1.5mm]
$(^{2}10_{\frac12} \to 10)_{_1}$&$\frac{N_c+21}{3}$ &$\beta_4$ &$-\frac{N_c+21}{3N_c}$ &$\frac{3(N_c+1)}{\sqrt{2}N_c}$
&$\frac{N_c+21}{2\sqrt{2}N_c}$ &0&$\beta_5$ &$\sqrt{\frac23}\frac{N_c+3}{N_c}$ &$\frac{N_c+21}{2\sqrt{6}N_c}$
 &$\frac{3-N_c}{4\sqrt{2}N_c}$ &$\frac{\sqrt{5}}{4}A_7$ \\[1.5mm]
$(^{2}10_{\frac12} \to 10)_{_2}$&$-\frac13$ &$\frac{1}{3N_c}$ &$\frac{1}{3N_c}$ &$-\frac{N_c+3}{6\sqrt{2}N_c}$
&$-\frac{1}{2\sqrt{2}N_c}$ &0&$\frac{N_c+7}{6\sqrt{15}N_c}$ &$-\frac{N_c+5}{10\sqrt{6}N_c}$ &$-\frac{1}{2\sqrt{6}N_c}$
 &$\frac{1}{60\sqrt{2}}$ &$\frac54A_8$ \\[1.5mm]
$^{2}10_{\frac32} \to 8$&1&$-\frac{1}{N_c}$ &$-\frac{4}{N_c}$ &$\frac{N_c+3}{2\sqrt{2}N_c}$ &0
&$-\frac{\sqrt{3}}{2N_c}$ &$\frac{2-N_c}{2\sqrt{15}N_c}$ &0&0&0&$-\frac{\sqrt{5}}{6}A_9$ \\[1.5mm]
$(^{2}10_{\frac32} \to 10)_{_1}$&$\frac{N_c+21}{3}$ &$\beta_4$ &$-\frac{N_c+21}{3N_c}$ &$\frac{3(N_c+1)}{\sqrt{2}N_c}$
&$\frac{N_c+21}{2\sqrt{2}N_c}$ &0&$\beta_5$ &$-\sqrt{\frac23}\frac{N_c+3}{N_c}$ &$-\frac{N_c+21}{2\sqrt{6}N_c}$
 &$\frac{N_c-3}{4\sqrt{2}N_c}$ &$\frac{\sqrt{5}}{4}A_7$ \\[1.5mm]
$(^{2}10_{\frac32} \to 10)_{_2}$&-1&$\frac{1}{N_c}$ &$\frac{1}{N_c}$ &$-\frac{N_c+3}{2\sqrt{2}N_c}$
&$-\frac{3}{2\sqrt{2}N_c}$ &0&$\frac{N_c+7}{2\sqrt{15}N_c}$ &$\sqrt{\frac32}\frac{N_c+5}{10N_c}$ &$\sqrt{\frac32}\frac{1}{2N_c}$
 &$-\frac{1}{20\sqrt{2}}$ &$\frac{5}{12}A_8$ \\[1.5mm]
$^{2}1_{\frac32} \to 8$&-1 &$\frac{1}{N_c}$ &0&$\frac{1}{\sqrt{2}N_c}$&0&$-\frac{N_c+3}{4\sqrt{3}N_c}$ &$-\frac{N_c+7}{4\sqrt{15}N_c}$ &0&0&0&
$-\frac{\sqrt{5}}{4}A_5$ \\[2mm] \hline
$\alpha_{n}$&$\sqrt{\frac{6}{7}}$ &$\frac 95\sqrt{\frac{3}{2}}$ &$\sqrt{\frac{3}{14}}$&$\frac{12}{5}\sqrt{3}$& $\frac 45\sqrt{3}$ &$6\sqrt{\frac{2}{7}}$&$\frac{18}{11}{\sqrt{5}}$&$\frac{12}{5}\sqrt{2}$&$ \frac{3}{\sqrt{8}}$&$3\sqrt{6}$& \\[2mm]
\hline \hline
\end{tabular}}
\end{center}
\end{table}

\begin{table}[!h]
\caption{Factors needed in Tables IV and V}
\begin{center}
\begin{tabular}{ccccccccc}
\hline \hline
$ \beta_1 $ && $\beta_2 $&& $\beta_3 $ && $\beta_4 $&& $\beta_5 $ \\
\hline
\vspace{-6.mm}
&&&&&&&&\\
$\frac{3-N_c(Nc+5)}{3N_c}$ &&
$-\frac{15+N_c(N_c+2)}{6\sqrt{15}N_c}$ &&
$-\frac{3+N_c(N_c+1)}{\sqrt{15}N_c}$ &&
$\frac{24+N_c(Nc+5)}{3N_c}$ &&
$\frac{(N_c-19)(Nc+3)}{6\sqrt{15}N_c}$ \\[2mm]
\hline
\vspace{2.5mm}
\end{tabular}
\begin{tabular}{ccccccccc}
\hline \hline
$A_1 $&&$A_2 $&&$A_3 $&&  $A_4 $&&  $A_5$\\
\hline
\vspace{-6.mm}
&&&&&&&&\\
$\frac{1}{N_c}\sqrt{\frac{(N_c+1)(N_c+3)(N_c+5)}{(N_c-1)}}$  &&
$\frac{1}{N_c}\sqrt{\frac{(N_c-1)}{(N_c+3)}}$ &&
$\frac{1}{N_c}\sqrt{\frac{(N_c+7)}{(N_c+3)}}$ &&
$\sqrt{\frac{(N_c+1)(N_c+5)}{N_c(N_c-1)}}$ &&
$\sqrt{\frac{(N_c-1)}{N_c}}$
\\[2mm] \hline
\vspace{2.5mm}
\end{tabular}
\begin{tabular}{ccccccc}
\hline \hline
$A_6 $&& $A_7 $&& $A_8 $&& $A_9$
\\
\hline
\vspace{-6.mm}
&&&&&&\\
$\sqrt{\frac{(N_c+7)}{N_c}}$ &&
$\sqrt{\frac{(N_c+5)}{N_c(45+N_c(N_c+6))}}$ &&
$\sqrt{\frac{(N_c+9)(N_c+1)(N_c-3)}{N_c(45+N_c(N_c+6))}}$ &&
$\sqrt{\frac{(N_c-1)(N_c+7)}{N_c(N_c+5)}}$\\[2mm]
\hline
\end{tabular}
\end{center}
\end{table}

The necessary isoscalar factors needed in the calculations of the actual matrix elements for the
 different channels are given in the Appendix.

It turns out that the sets of operators listed in Tables IV and V are linearly dependent
when restricted, for arbitrary $N_c$, to the $SU(3)$  multiplets corresponding to the ones in the $\bf 70$-plet (according to Table XV of the Appendix). For the $SU(3)$ preserving operators one finds the following
set of bases after eliminating linear dependencies, namely,  for the S-wave decays the LO basis is
given by $ \{ {\cal B}_1,\;{\cal B}_2,\;{\cal B}_6\}$ and the NLO basis is given by
$ \{ {\cal B}_1,\;{\cal B}_2,\;{\cal B}_3,\;{\cal B}_4,\;{\cal B}_6\}$ and in addition the
$SU(3)$ breaking operators $ \{ {\cal B}^{SB}_1,\;{\cal B}^{SB}_2,\;{\cal B}^{SB}_3\}$. For
the D-wave decays the LO basis is given by  $ \{ {\cal B}_1,\;{\cal B}_2,\;{\cal B}_6,\;{\cal B}_8\}$
and the NLO basis is given by
$ \{ {\cal B}_1,\;{\cal B}_2,\;{\cal B}_3,\;{\cal B}_5,\;{\cal B}_6,\;{\cal B}_8,\;{\cal B}_9\}$ and
the  $SU(3)$ breaking operators $ \{ {\cal B}^{SB}_1,\;{\cal B}^{SB}_2,\;{\cal B}^{SB}_3\}$. For the
available data inputs to the fits one finds that one of the D-wave $SU(3)$ breaking operators can be
eliminated due to linear dependency. We will choose in this case to eliminate the third one. The
normalization factors for the symmetry breaking operators are as follows: for the S-wave operators
they are $\alpha^{SB}_1=\frac{\sqrt{3}}{2} $,   $\alpha^{SB}_2= \sqrt{\frac 23}$ and
 $\alpha^{SB}_3= \sqrt{\frac 65}$, and for the two D-wave operators we use in the fit,
 $\alpha^{SB}_1=\frac 23\,\sqrt{\frac{10}{21}} $ and
 $\alpha^{SB}_2=\sqrt{\frac{3}{10}} $.

 One comment concerning the mixing of states. In the $SU(3)$
 symmetry   and  strict large $N_c$ limits the two mixing angles
 have definite values independent of coefficients, namely
 $\theta_1=\arccos\left(\frac{1}{\sqrt{3}}\right)$ and $\theta_3=\arccos\left(
 \frac{1}{\sqrt{6}}\right)$ \cite{PirjolSchat}. 
 It was found
 \cite{GSS70decays} that to leading order in $1/N_c$ with these
 mixing angles the following transitions vanish: the S-wave
 $N(1535)\to \pi N$, $N(1650)\to \eta N$, and the D-wave
 $N(1700)\to \pi N$ and $N(1520)\to \eta N$. We have checked that
 indeed these cancellations take place. Another channel suppressed
 in those limits is $\Lambda(1700)\to\pi \Sigma$. In reality the
 mixing  angles differ significantly from the strict large $N_c$
 limit angles, primarily because the subleading in $1/N_c$
 hyperfine interaction contributes much more to the masses than
 the ${\cal{O}}(N_c^0)$ operators (spin-orbit type interactions),
 which in the strict large $N_c$ limit  give the limit angles.
 The deviations from those cancellations  in the decay amplitudes are thus explained
 primarily by the mixings, but also by subleading effects in
 $1/N_c$.

It is useful to asses the potential predictive power of the   analysis presented here.
The number of possible  different decay amplitudes  in the isospin limit is as follows: 17   in $\bf 8 \to {\bf 8}$,  12  in $\bf 8 \to {\bf 10}$ or ${\bf 10} \to {\bf 8}$, 13  in ${\bf 10} \to {\bf 10}$, and 4   in ${\bf 1} \to {\bf 8}$. Thus, there are 87 S-wave   and 153 D-wave    amplitudes. 
Using the empirical baryon and meson masses together with the $1/N_c$ predictions of Ref.\cite{SGS} for those in the 70-plet
which are still unknown,
we estimate that the number of phase space allowed decays are   52  for S-waves and  94 for D-waves.
For S-wave
transitions, there are a total of seven possible $R_{B^*} \to R_{B}$ transitions, which
is one more than the number of $SU(3)$ preserving S-wave
operators at NLO.  This implies one  overall relation between amplitudes in $SU(3)$ symmetric
limit.  
In the isospin limit we have 87 channels, and since at NLO 
  we have included eight operators, there are in principle 79 relations.  For the D-waves there are  twelve $R_{B^*} \to R_{B}$ transitions,
 which in $SU(3)$ symmetry limit with the ten operators leave two relations. The number of  different
 channels in the isospin limit is equal to 153, and thus, with the eleven operators we include, there
 are in principle 143 relations between amplitudes.
   Most of the  relations would represent a test of $SU(3)$ symmetry and its breaking by the   1-body
   operators, as one may expect.  With the available information on partial widths we can  consider a limited number of
   relations at leading order, which we discuss below.

\section{Analysis of the Partial Widths}

Only a fraction of the possible partial decay channels of the
negative parity baryons has been empirically determined \cite{PDG}.
This relatively poor database imposes significant  limitations in the
accuracy and robustness of the conclusions of the present
analysis, as we discuss later. Of the 87 S-wave  and 153 D-wave
different partial widths which would be possible in the $[{\bf
70},1^-]$ decays (respectively  52 and  94 if phase space is taken into account), only 16 S-wave and 25 D-wave partial widths are
known, a few of them only as bounds.

A very important consistency check of the viability of the $1/N_c$
analysis is given in terms of coefficient independent relations,
namely relations independent of the values of the coefficients
$C_n$. These relations exist at each order in the expansion, and
can be given at the level of reduced partial decay widths defined below. These
relations  provide in principle  predictions for unknown partial
decay withs  accurate to the order of the expansion. Since the
widths are quadratic in the fitting coefficients, the total number
of coefficient independent relations is  $N_{\rm rel}=N_{\rm
ch}-N_{\rm par}(N_{\rm par}+1)/2$, where $N_{\rm rel}$ is the
number of relations,  $N_{\rm ch}$ is the number of  partial
widths, and  $N_{\rm par}$ the number of coefficients $C_n$.
With
the numbers of channels mentioned above, and the number   of
operators involved  at LO in $1/N_c$, where the $SU(3)$ breaking
effects are relegated to sub-leading order, one finds  81
relations for S-wave decays and 143  for D-wave decays. A smaller number results because of channels suppressed by phase space, and  a much
smaller number yet  for the actually known decays. Various
such relations in the form of ratios of reduced widths are given
below.  Some of those can be tested  and some give LO predictions
for mixing angles. We do not discuss NLO relations; in principle there are 51 for S-waves and 108 for D-waves, not taking into account phase space suppressed channels.

We discuss now some reduced width ratios which can be tested with the known partial decay widths. The list is not exhaustive.
The reduced partial decay  widths are defined by
$\tilde\Gamma=\Gamma/f_{phs}$, where $f_{phs}=\frac{k_P}{8
\pi^2}\frac{M_B}{M_{B*}} \left(
\frac{k_P}{\Lambda}\right)^{2\ell_P}$.  These are basically
squares of decay amplitudes with    the centrifugal factor  removed. For
the S-wave decays we obtain the ratios shown in Table VII. These
results show  that there is a significant discrepancy in the
predictions for the mixing angle $\theta_1$ resulting from
non-strange versus strange decay channels. The mixing angles get
$SU(3)$ breaking corrections which are indeed important as we will
find in the NLO results. The actual ratio $\frac{\tilde{\Gamma}(
N(1650)\to \pi N)}{\tilde{\Gamma}( \Sigma(1750)\to \eta\Sigma)}$
is very different from the LO prediction. This indicates large
$SU(3)$ breaking. Indeed, one expects $SU(3)$ breaking to be a
30\% effect at the level of amplitudes or a 60\% effect at the
level of the reduced widths. In this case the deviation  is
significantly larger than that.

\begin{table}[!h]
\caption{S-wave decays: coefficient independent ratios of reduced widths at LO.}
\begin{center}
\begin{tabular}{lcccccl}
\hline \hline
Ratio &&  LO Ratio& &Empirical &&~~~~ Prediction \\[-5mm]\\
\hline
\vspace{-6.mm}
 &&&&&&\\
 $\frac{\tilde{\Gamma}(N (1535)\to \pi N)}{\tilde{\Gamma}(N (1650)\to \pi N)}$              &&  $\left( \frac{\sqrt{2} \cos\theta_1-\sin\theta_1}{\cos\theta_1+\sqrt{2}\sin\theta_1}\right)^2$          &&   $0.6\pm0.2$             &&$\theta_1=\left\{\begin{array}{l}
 0.3\pm0.1\\ 1.6\pm0.1
 \end{array}\right.$\\
 $\frac{\tilde{\Gamma}( \Delta(1620)\to \pi N)}{\tilde{\Gamma}(\Delta (1700)\to \pi \Delta)}$   &  &2/5          && $0.30\pm0.15$   &            &\\
 $\frac{\tilde{\Gamma}(N (1535)\to \pi N)}{\tilde{\Gamma}(\Delta (1620)\to \pi N)}$             & & $3+\cos 2\theta_1-\sqrt{8}\sin 2\theta_1$       &    &         $2.1\pm 0.7$   &    &$\theta_1=\left\{\begin{array}{l}
 0.3\pm0.1\\ 1.6\pm0.1
 \end{array}\right.$  \\
 $\frac{\tilde{\Gamma}(N (1535)\to \eta N)}{\tilde{\Gamma}(N (1650)\to \eta N)}$            &&  $\left( \frac{ \cos\theta_1+\sqrt{2}\sin\theta_1}{\sqrt{2}\cos\theta_1-\sin\theta_1}\right)^2$      & &     $13\pm 11$           &&$\theta_1=\left\{\begin{array}{l}
 0.7^{+0.1}_{-0.3}\\ 1.6^{+0.3}_{-0.1}
 \end{array}\right.$\\
 $\frac{\tilde{\Gamma}(N (1520)\to \pi \Delta)}{\tilde{\Gamma}( \Delta(1620)\to \pi N)}$        &&  $3-2 \cos 2\theta_3+\sqrt{5}\sin 2 \theta_3$         & & $0.44\pm0.20$             &  &  $\theta_3=\left\{\begin{array}{l}
 2.4\pm0.1\\ 2.9\pm0.1
 \end{array}\right.$\\
 $\frac{\tilde{\Gamma}(N (1650)\to \pi N)}{\tilde{\Gamma}(\Lambda (1670)\to \pi\Sigma)}$    &&   $\left( \frac{ \cos\theta_1+\sqrt{2}\sin\theta_1}{\sqrt{2}\cos\theta_1-\sin\theta_1}\right)^2$          & &   $7.6\pm 4.3$           & &     $\theta_1=\left\{\begin{array}{l}
 0.6^{+0.1}_{-0.2}\\ 1.3^{+0.2}_{-0.1}
 \end{array}\right.$\\
 $\frac{\tilde{\Gamma}( N(1650)\to \pi N)}{\tilde{\Gamma}( \Sigma(1750)\to \eta\Sigma)}$        & & 1           &&     $0.12\pm     0.12$     & &\\[2mm]
\hline
\vspace{2.5mm}
\end{tabular}
\end{center}
\end{table}

For the D-wave decays most relations we can derive involve the mixing angles.
Those not involving mixing angles  
 and which can be tested  are shown in Table VIII.  We have looked at
relations that can be tested when only the 1-body operator is
kept. Algebraically, this corresponds to a quark model with
emission of a meson coupled to the excited quark, in the $SU(6)$
limit. For instance, the relations that can predict the mixing
angles give angles very different from those obtained in the fit.
In fact, we will see that a 2-body LO operator is important for
the D-wave  fits.

\begin{table}[!h]
\caption{D-wave decays: coefficient independent ratios of reduced widths at LO.}
\begin{center}
\begin{tabular}{lcccl}
\hline \hline
Ratio &&  LO Ratio& &Empirical  \\[-5mm]\\
\hline
\vspace{-6.mm}
 &&&&\\
 $\frac{\tilde{\Gamma}(\Lambda(1830)\to \pi \Sigma)}{\tilde{\Gamma}(\Sigma (1775)\to \pi \Lambda)}$                 &&  3              &&   $2.7\pm 1.4$     \\[2mm]
 $\frac{\tilde{\Gamma}( N(1520)\to \pi N)}{\tilde{\Gamma}(\Lambda(1690)\to \pi \Sigma)}$    &  & 1          && $0.5\pm 0.2$   \\[2mm]
 $\frac{\tilde{\Gamma}(N (1520)\to \pi N)}{\tilde{\Gamma}(\Sigma(1670)\to \pi \Lambda)}$            & & 3      &    &         $2.2\pm 1.4$   \\[2mm]
 $\frac{\tilde{\Gamma}(\Sigma(1670)\to \pi \Lambda)}{\tilde{\Gamma}(\Sigma(1670)\to \pi \Sigma)}$           && 1/2  & &     $0.12\pm0.10$           \\[2mm]
 $\frac{\tilde{\Gamma}(N (1675)\to \pi N)}{\tilde{\Gamma}( \Lambda(1830)\to \pi \Sigma)}$       && 1       & & $0.9\pm0.4$               \\[2mm]
 $\frac{\tilde{\Gamma}(N (1675)\to \pi N)}{\tilde{\Gamma}(\Sigma (1775)\to \pi\Lambda)}$    &&   3      & &   $2.3\pm 0.6$              \\[2mm]
$\frac{\tilde{\Gamma}(N (1675)\to \pi N)}{\tilde{\Gamma}(\Sigma (1775)\to \pi\Sigma)}$  &&   3/2      & &   $7\pm 4$                \\[2mm]
 $\frac{\tilde{\Gamma}( \Sigma(1775)\to \pi \Sigma)}{\tilde{\Gamma}( \Sigma(1775)\to \pi\Sigma^*)}$         & & 8/7           &&     $0.06\pm     0.03$     \\[2mm]
 $\frac{\tilde{\Gamma}(\Lambda(1690)\to \pi \Sigma)}{\tilde{\Gamma}(\Sigma (1670)\to \pi\Lambda)}$  &&   3      & &   $4.6\pm 3.2$           \\[2mm]
\hline
\vspace{2.5mm}
\end{tabular}
\end{center}
\end{table}

Now we proceed to discuss the fits. The coefficients and mixing angles obtained in the fits are depicted in Table IX. Tables  X through XIV show the fitted widths as well as the predictions.
The LO fit is performed using the operator bases $\{{\cal{B}}_1, ~{\cal{B}}_2,~{\cal{B}}_6\}$ for the S-wave amplitudes and
 $\{{\cal{B}}_1, ~{\cal{B}}_2,~{\cal{B}}_6,~{\cal{B}}_8\}$ for the D-wave amplitudes. No $SU(3)$ breaking effects are therefore included. The  matrix elements are not expanded  in $1/N_c$. Both S- and D-wave partial widths are fitted simultaneously since both depend on the mixing angles $\theta_1$ and $\theta_3$, which are fitted. As a check we performed a fit of the non-strange sector of decays, with similar results to those obtained in previous work \cite{GSS70decays} (we note that the input partial widths have slightly changed from that analysis to the present one).  In particular, the main problem with that fit is the discrepancy in the $N(1535)\to  \eta N$.  When all available channels are included, the $\chi^2_{\rm dof}$ is quite large, predominantly due to the difficulty in describing the $\eta$ channels as well as several of the excited hyperon decays.  For instance, if one would like to have the mixing angles $\theta_{1,\;3}$ similar to the ones from the non-strange analysis, one has to remove the channels $\Lambda(1670)\to \overline{K} N$ and $\Lambda(1690)\to \overline{K} N$. In fact, in quark models these channels had been found long ago to be problematic to describe \cite{FKR}.  The issue of large $SU(3)$ breaking effects is clearly manifested by the shortcomings of the LO fit, as well as the LO ratios we discussed earlier. In particular we find that the result for the  S-wave partial width $\Delta(1700)\to \pi \Delta$ is rather sensitive to the removal of the mentioned $\Lambda$ channels, and others are also  significantly affected. After analyzing the NLO fit, one realizes that indeed the mentioned $\Lambda$ channels and the $\eta$ channels cannot be described consistently at LO. This is based on the stability of the coefficients of the dominant operators in the LO fit and of the mixing angles. One learns from the LO fit that the dominant operators are for the S-wave decays the 1-body ${\cal{B}}_1$  and important but with smaller contributions  the 2-body  operator ${\cal{B}}_6$, and for the D-wave decays the  ${\cal{B}}_1$  and  the 2-body  ${\cal{B}}_8$ operators.

 The NLO fit is complicated by the large number of coefficients
 and angles, in total 23 parameters. Since the number of inputs is
 at most 41, there are multiple minima of the $\chi^2$ function
 which have similar values. The criterion is imposed that the
 coefficients of dominant operators at LO and the mixing angles
 which are not affected by $SU(3)$ breaking effects, namely
 $\theta_{N_1}$ and  $\theta_{N_3}$,  deviate from the LO results as expected by
 NLO corrections. A fit satisfying that criterion is obtained.
 This fit shows that the coefficients of the $SU(3)$ preserving
 operators are of  natural size, and most of them are actually
 smaller than natural size. On the other hand the $SU(3)$ breaking
 S-wave operator  ${\cal{B}}^{SB}_1$  and the D-wave operator
 ${\cal{B}}^{SB}_2$ have coefficients roughly a factor two larger 
 than natural size. ${\cal{B}}^{SB}_1$  only contributes to $\eta$
 channels, and eliminating it increases the $\chi^2_{\rm dof}$
 from 1.5 to 1.9, and at the same time the coefficients of the S
 wave operators ${\cal{B}}^{SB}_{2,\;3}$ become unnaturally large.
 This exercise  indicates that there is significant correlation
 among the coefficients of the S-wave $SU(3)$ breaking operators.
 Evidently, the $SU(3)$ breaking operators are crucial for
 describing  the $\eta$ channels. One also finds that $SU(3)$
 breaking is significant in the mixing angles; in particular it is
 unnaturally large for the $\Lambda$ baryons.
 
The analysis leads to the following observations.
\begin{itemize}
\item The 1-body operators  ${\cal{B}}_1$ are dominant in S- and
  D-waves. This supports the quark model picture in which the
  meson is predominantly emitted from the excited quark. 
\item  2-body operators are less important but necessary to obtain good
  fits. These operators will in particular encode the longer range
  dynamics of the decays. At LO they have smaller or much smaller
  coefficients than  the 1-body operator  ${\cal{B}}_1$, and at NLO
  their coefficients are in general smaller than the natural size.
  Note that this conclusion is under the criterion of selecting
  the fits which have LO coefficients stable as one moves to the
  NLO fit. Thus, one can conclude that NLO fits consistent with a
  $1/N_c$ power counting are possible. We also note that such fits
  are indeed the ones with lowest $\chi^2$ we have found.
  \item As Table IX shows, several operators carry coefficients consistent with zero within error. One can eliminate those operators and perform a NLO fit where the coefficients of the relevant  operators 
  do not change significantly,  and  $\chi^2_{\rm dof}\sim 1.2$.
  \item  A lower $\chi^2_{\rm dof}\sim 1$  can be obtained by reducing by a factor $\sim 0.6$  the exponent of  the centrifugal barrier. This however does not  give any significant change to the  outcome of the analysis.
 \item For the nucleon's mixing angles,  previous analysis of the non-strange sector gave $\theta_{N_1}=0.39\pm 0.11$ and $\theta_{N_3}=2.82$ or $2.38$$\pm 0.11$,
  to be compared respectively with  $0.42\pm 0.07$ and $2.74$ or  $2.36 \pm 0.09$ obtained in the present analysis.  
  We note that the ambiguity in $\theta_{N_3}$ found in the previous analysis persists in the present analysis at both LO and NLO level.  
  A similar ambiguity is found at NLO for the mixing angle $\theta_{\Sigma 1}$. For the nucleons the ambiguity can be sorted out by analyzing the photocouplings \cite{GMS,GdeUS}
 \item The $SU(3)$ breaking effects are of unnaturally large
  magnitude by 
  roughly a factor two. In particular they manifest
  themselves in the $\eta$ channels, where the LO fit gives very
  poor description. This problem had been noticed when those
  channels were included in the analysis of the non-strange decays
  \cite{GSS70decays}. The very important S-wave channel
  $N(1535)\to \eta N$ is too small at LO by a factor two, while
  the small S-wave $N(1650)\to \eta N$ is also under-predicted at
  LO by a factor four. On the other hand the $\pi N$ channels are
  well described at LO for both resonances.
 \item The S-wave  decay $\Lambda(1405)\to \pi \Sigma$  is well
  described in all fits. It is sensitive to the presence of the
  2-body operator ${\cal{B}}_6$. On the other hand the D-wave
  decay  $\Lambda(1520)\to \pi \Sigma$ is well described while the
  $\Lambda(1520)\to \overline{K} N$ is poorly described at LO. A
  clear example of $SU(3)$ breaking effects.
  \item The decays  $\Lambda(1670)\to   \overline{K} N$ (S-wave) and
  $\Lambda(1690)\to   \overline{K} N$ (D-wave) are poorly described at LO
  if one requires that the mixing angles $\theta_{N_{1,\;3}}$ are
  similar to the values obtained in fits of the non-strange decays
  only or the NLO fits. At NLO these decays are improved   because
  of the $SU(3)$ breaking in the mixing angles in particular.
 \item The S-wave decay $\Delta(1700)\to \pi \Delta$ is particularly
  sensitive to the 2-body operator  ${\cal{B}}_2$, and found to be
  very sensitive to the inclusion or not of the latter $\Lambda$
  channels in the LO fit.
 \item The NLO results show that the mixing angles are strongly
   affected by $SU(3)$ breaking effects. To obtain a more accurate
   picture it will be necessary to carry out an analysis of masses
   and photo-couplings along with the decays \cite{nextpaper}.
 \item We provide predictions for the unknown channels of known
  {\bf 70}-plet states, and here we discuss some of them.  Little
  can be concluded from predictions of small partial widths,
  except that the corresponding channels will be most likely
  experimentally inaccessible. On the other hand several large
  partial widths are predicted, which require some discussion.
  \begin{itemize}
    \item   The $N(1700)$ is given in the PDG with three stars, but its existence is challenged by several recent analyses \cite{GW,EBAC,Juelich}, while other analyses confirm it   \cite{Manley-Thoma}.  Since the photo-couplings and also electro-couplings of the $p(1700)$ and $n(1700)$  are small \cite{PDG,GMS,Mokeev}, the access to   $N(1700)$  by these means is limited.
  In our fits we included the PDG estimate for the $\pi N$ channel, but disregarding this  input gives no significant change to the fits. The S-wave partial width of 
  $N(1700)\to \pi \Delta$ is predicted to be the largest one   while the D-wave channel has a small  width. This contradicts partial wave analyses \cite{Manley-Thoma}  where it is claimed that the D-wave    is significantly larger than the S-wave.  According to the PDG \cite{PDG}, the total width of the   $N(1700)$ is rather uncertain, and an estimate of the  $N(1700)\to \pi \Delta$  partial width from the $\pi NN$ width is viable provided one assumes it to proceed mostly through the $\pi \Delta$ S-wave channel if one is to obtain  a reasonable fit at NLO. This however affects the fit significantly.  We have chosen not to include  as input the estimate of the   $N(1700)\to \pi \Delta$ S-wave partial width, and this remains therefore an open problem   requiring  further empirical progress.
 \item A similar problem to
  that with the   $N(1700)$ is found for the $\Lambda(1690)\to \pi
  \Sigma^*$, where the empirical estimate for this partial width,
  assumed to  be S-wave and estimated from the dominance of this
  channel in the decays $\pi\pi \Lambda$ and $\pi \pi \Sigma$, is
  around $20$ MeV, while the fits predict it to be  larger than $100$
  MeV. Giving that estimate as input has similar effects as the
  input of     $N(1700)\to \pi \Delta$. We do not include this
input either.
  \item Briefly, the inclusion of these S-wave inputs lead to the following: the inclusion of either estimate leads to the same values of the coefficients within errors, 
  the main changes with respect to not including either of them  being the values of the S-wave coefficients,  which in an indirect way alter significantly the angle $\theta_{N_1}$. 
  \item The prediction for the D-wave partial width $N(1675)\to \pi \Delta$ of $75\pm 22$ MeV is in good agreement with the estimate based on the total width and the other significant partial width into $\pi N$.
  \item The S-wave partial widths of the $\Lambda(1800)$ to $\eta \Lambda$ and to $\pi \Sigma$ are predicted to be  approximately equal to the empirical one for the channel $\overline{K}N$, and in agreement with the total width.
  \item The D-wave partial width $\Lambda(1830)\to \pi \Sigma^*$ is predicted to be the largest one, and it agrees with empirical lower bound of 15\% for the branching ratio.   The total predicted width is $116\pm 20$ MeV, to be compared with the PDG   \cite{PDG} estimate of 60 to 110 MeV.
  \item For the D-wave $\Xi(1820)$ the PDG   \cite{PDG} gives a total width $\sim 25\pm 15$ MeV  \cite{PDG}. The predictions  given in Table XIII are dependent on the choice of the mixing angle $\theta_{\Xi3}$  which cannot be determined by the current analysis. The predictions are given for the choice  $\theta_{\Xi3}=\theta_{N_3}=2.74$. The dominant channels are  the $\overline{K} \Sigma$ and $\overline{K} \Lambda$.  The total width predicted is about a factor two larger than the one from the PDG.
If one requires that the total width  be reproduced, the mixing angle becomes  $\theta_{\Xi3}\sim 1.9$.    \end{itemize}
 \item Several channels where there can be $SU(3)$ symmetry breaking effects are predicted to have significant partial widths, and could therefore become empirically accessible. Such channels are the S-wave $\Lambda(1690)\to \pi \Sigma^*$, $\Lambda(1800)\to\eta \Lambda$,  $\Lambda(1800)\to\pi \Sigma$, $\Sigma(1670)\to \pi \Sigma^*$, and   $\Sigma(1750)\to \pi \Lambda$, and the D-wave $N(1675)\to\eta N$, $\Lambda(1830)\to \eta \Lambda$,    $\Lambda(1830)\to \pi \Sigma^*$,
 $\Sigma(1670)\to \pi \Sigma^*$ and $\Sigma(1775)\to \overline{K} \Delta$. All these channels involving $\pi$ meson are not affected by the 1-body $SU(3)$ breaking effects, and would give additional information  on breaking at the 2-body level.  
\end{itemize}

\begin{table}[!h]
\caption[Fit parameters.]{Parameters from LO and NLO fits. Note ambiguities in some angles; the rest of the parameters differ within the errors, therefore parameters have been given for the first quoted angle in the ambiguous cases.}
\begin{center}
\begin{tabular}{ccccllccclclcccllccc}
\hline \hline
S-wave coeff &&  LO && {NLO}&&   D-wave coeff  &&  LO && {NLO} \\
\hline
$c_{S1}$ &&20.6(1.0) && 19.6(1.6)&&$c_{D1}$ && 2.79(0.12)&& 2.59(0.20)&&&&&&\\
$c_{S2}$ && 0.92(1.2)&&$-3.1$(1.2)&&$c_{D2}$ && 0.03(0.08)&&$-0.28$(0.09)&&&&&&\\
$c_{S3}$ && 0&&5.2(5.8)&&$c_{D3}$ && 0&& $0.92(1.0)$&&&&&&\\
$c_{S4}$ && 0&&9.2(4.0) && $c_{D5}$ && 0 && $1.54(0.39)$ &&&&&&\\
$c_{S6}$ &&$ -6.36(1.2)$&& $-8.2$(1.8)&& $c_{D6}$ && 0.03(0.17) && $-0.26(0.26)$ &&&&&&\\
$b_{S1}$ && 0 && 40.3(14)&& $c_{D8}$ && 1.18(0.12) && $1.28(0.15)$ &&&&&&\\
$b_{S2}$ && 0 && $-0.25$(14)&&$c_{D9}$ && 0 && $0.09(0.85)$ && && && \\
$b_{S3}$ && 0 && $-3.0$(14)&&$b_{D1}$ && 0 && $1.0(1.0)$ && &&\\
&& && && $b_{D2}$ && 0 && $5.6(0.5)$ && &&\\
\hline \hline
Angle && LO && {NLO} && && LO && {NLO} \\ \hline
$\theta_{{N_1}}$ && 0.58(0.06)&& 0.42(0.07)&& $\chi^2_{\rm dof}$ && 4.8 && 1.52\\
$\theta_{{N_3}}$ &&
$\left\{\begin{array}{c}2.82(0.05)\\
2.36(0.06)
 \end{array}\right.$
&& $ \left\{\begin{array}{c}2.74(0.09)\\
2.36(0.10)
 \end{array}\right. $ && $\rm dof$ && 30 && 18
\\
$\theta_{{\Lambda_1}}$ &&$\theta_{{N_1}}$ && 0.99(0.09)&& && && && && \\
$\theta_{{\Lambda_3}}$ && $\theta_{{N_3}}$ && 1.56(0.07)&&&&&&&& &&\\
$\theta_{{\Sigma_1}}$ && $\theta_{{N_1}}$&& $\left\{\begin{array}{c}0.28(0.16)\\
3.03(0.17)
 \end{array}\right. $ &&&&&&&&&&&&\\
$\theta_{{\Sigma_3}}$ &&$\theta_{{N_3}}$ &&2.19(0.47)&&&&&&&&&&&&\\
\hline \hline
\end{tabular}
\label{ngttab6}
\end{center}
\end{table}

\clearpage
\begin{table}[!h]
\caption{The decay widths of $N$ states in ${\bf 70}$-plet
whose mass is currently experimentally known. Values are in MeV.}
\begin{center}
\begin{small}
\begin{tabular}{l ccc ccc ccc}
\hline \hline
& \multicolumn{3}{c}{$N(1535)$}&
     &\multicolumn{4}{c}{$N(1520)$}
\\
\cline{2-4}\cline{6-9}
 & $\pi N$ &$\eta N$ &$\pi \Delta$ &\hspace{2mm}\  &\multicolumn{2}{c}{$\pi \Delta$} &$\pi N$ &$\eta N$
\\
\cline{2-4}\cline{6-9}
PW    &S          &S          &D          && S         &D         &D         &D          \\
LO &57(17)     &33(6) & 0.3(0.2)  &&    8.9(4.3)   &  8.1(1.0)&77(7)   & 0.09(0.01)     \\
NLO &57(19)     &73(44) & 0.9(0.7)  &&    9(11)   &  10(2)&72(11)   & 0.26(0.07)     \\
Exp     &68(19)     & 79(17)    &0.8(0.8  ) && 9.6(4.1)  &13.6(2.7) &69(10)    &0.26(0.05) \\
\hline \hline \vspace{2.5mm}
\end{tabular}
\begin{tabular}{l cccc c cccccc}
\hline \hline
     & \multicolumn{4}{c}{$N(1650)$}&
     & \multicolumn{6}{c}{$N(1700)$}
\\
\cline{2-5}\cline{7-12}
 & $\pi N$ &  $\eta N$ & $K \Lambda$&  $\pi \Delta$ && \multicolumn{2}{c}{$\pi \Delta$} & $\pi N$ & $\eta N$ &  $K \Lambda$& $K \Sigma$
\\
\cline{2-5}\cline{7-12}
PW    &S          &S          & S         &D         &&S       &D   &D          &D        &D         &D     \\
LO &143(26)    &2.5(1.6)    &9.8(2.9)   &4.8(2.6)   &&215(57)  &2.9(2.4) &11.4(8.5)      &0.52(0.25) & 0.13(0.08)      &$\sim 0$ \\
NLO  &133(33)     &12.5(11.0)     &11.5(6.4)     &5.1(5.8)   &&297(111)     &0.3(2.0) &12(13)   & $\leq 0.15$&$\leq 0.03 $      &$\sim 0$   \\
 Exp     &128(33)    &10.7(5.9)  &11.5(6.7)  &6.6(5)    &&        &    & 10(7)     &         &1.5(1.5)  &      \\
\hline \hline
\vspace{2.5mm}
\end{tabular}
\begin{tabular}{l ccc cc}
\hline \hline
    & \multicolumn{4}{c}{$N(1675)$} \\
\cline{2-5}
    & $\pi N$ & $\eta N$ &$K \Lambda$&$\pi \Delta$ \\
\cline{2-5}
PW    &   D       &      D    &   D         & D      \\
LO &52(8)    &2.6(0.4)    & 0.02(0.01)      & 72(9)  \\
NLO &51(12)   &6.3(2.5)     &$\leq 0.1$       &75(24)  \\
Exp     &  59(10)   &           & 0.75(0.75)  &        \\
\hline \hline
\end{tabular}
\label{ngttab8}
\end{small}
\end{center}
\end{table}
\begin{table}[!h]
\caption{The decay widths of $\Lambda$ states in ${\bf 70}$-plet
whose mass is currently experimentally known. Values are in MeV.}
\begin{center}
\begin{tabular}{l ccc ccc ccccc}
\hline \hline
& \multicolumn{4}{c}{$\Lambda(1670)$}&
     &\multicolumn{5}{c}{$\Lambda(1690)$}
\\
\cline{2-5}\cline{7-11}
 & $\overline{K} N$ &$\eta \Lambda$ &$\pi \Sigma$ &$\pi \Sigma^*$ &
    &\multicolumn{2}{c}{$\pi \Sigma^*$}&$\overline{K} N$ & $\eta \Lambda$ &$\pi \Sigma$
\\
\cline{2-5}\cline{7-11}
PW    &S          &S          &S            &D      &&S           &D          &D         &D        & D        \\
LO &113(24)     &0.11(0.12)   &1.8(2.0)     &0.16(0.09) &&7.3(3.5)     &9(1)     &60(6)  &$\sim 0$     &9.0(0.9)    \\
NLO &9(15)     &6.1(4.3)     & 15(11)       &0.04(0.10)  &&114(49)        &2.1(1.5)       &16(5)      &$\sim 0$     & 5.3(2.9)      \\
Exp     & 9.4(3.6)  &6.6(3.6)   &15(7.5)      &       &&            &           &  15(4)   &         &  18(6.7) \\
\hline \hline
\vspace{2.5mm}
\end{tabular}
\begin{tabular}{l ccc ccc cccc}
\hline \hline
& \multicolumn{4}{c}{$\Lambda(1800)$}&
     &\multicolumn{5}{c}{$\Lambda(1830)$}
\\
\cline{2-5}\cline{7-11}
 & $\overline{K} N$ &$\eta \Lambda$ &$\pi \Sigma$ &$\pi \Sigma^*$ &
    &$\overline{K} N$ &$\eta \Lambda$ &$\pi \Sigma$&$K \Xi$&$\pi \Sigma^*$
\\
\cline{2-5}\cline{7-11}
PW      &S           &S          &S         &D      &&D           &D          &D         &D     &D      \\
LO   &43(13)&30(4)     &150(20)     &3.0(1.6)   &&3.0(1.6)       &3.5(0.3)     &69(6)    &$\sim 0$&54(7)    \\
NLO   &100(73)       &94(47)       &109(25)       & 5.9(5.2)  && 12(4)        & 9.6(2.5)       & 38(11)       & $\sim 0$  & 57(18)   \\
Exp       &98(40)      &           &          &       &&5.5(3.4)    &           &46.7(22)  &      &       \\
\hline \hline
\vspace{2.5mm}
\end{tabular}
\begin{tabular}{l c ccc}
\hline \hline & $\Lambda(1405)$ &
&\multicolumn{2}{c}{$\Lambda(1520)$}
\\
\cline{2-2}\cline{4-5}
    &$\pi \Sigma$&
    &$\overline{K} N$&$\pi \Sigma$
\\
\cline{2-2}\cline{4-5}
PW     &S            && D      &D         \\
LO  &50(19)         && 2.7(0.4)   &8.2(1.3)  \\
NLO  &50(9)         && 6.7(1.1)    & 6.9(1.8)      \\
Exp      &50(5)        && 7(0.5) & 6.5(0.5) \\
\hline \hline
\end{tabular}
\label{ngttab9}
\end{center}
\end{table}
\clearpage

\begin{table}[!h]
\caption{The decay widths of $\Sigma$ states in ${\bf 70}$-plet
whose mass is currently experimentally known. Values are in MeV.}
\begin{center}
\begin{tabular}{l ccc cc}
\hline \hline
     &\multicolumn{5}{c}{$\Sigma(1670)$}
\\
\cline{2-6}
    &\multicolumn{2}{c}{$\pi \Sigma^*$} &$\overline{K} N$ & $\pi \Lambda$ &$\pi \Sigma$ \\
\cline{2-6}
PW    &S          &D        &D        &D         &D        \\
LO &1.5(0.7)  & 1.5(0.2)   &2.1(0.5)   &4.8(0.5)     &46(5)    \\
NLO &4(11)       & 1.5(0.9)    & 2.5(1.4)     & 7.0(2.9)      & 28(11)     \\
Exp     &           &         &6(2.7)   &6(3.6)    &27(12.7) \\
\hline \hline
\vspace{2.5mm}
\end{tabular}
\begin{tabular}{l ccc ccc}
\hline \hline
& \multicolumn{6}{c}{$\Sigma(1750)$}
\\
\cline{2-7}
 & $\overline{K} N$ &$\pi \Lambda$ &$\pi \Sigma$  &$\eta \Sigma$ &$\overline{K} \Delta$ &$\pi \Sigma^*$ \\
\cline{2-7}
PW     &S          &S           &S         &S          &D        &   D    \\
LO  &45(8)      &51(7)       &6.2(5.3)    &14(2) &0.07(0.04)   &0.5(0.3)   \\
NLO  &30(34)  & 38(12)       &4.2(7.6)      &53(28)      & 0.4(0.2)    & 0.4(0.5)    \\
Exp      &27.5(21)   &            &4.4(4.4)  &38.5(28)   &         &        \\
\hline \hline
\vspace{2.5mm}
\end{tabular}
\begin{tabular}{l ccc ccc}
\hline \hline
& \multicolumn{6}{c}{$\Sigma(1775)$}
\\ \cline{2-7}
 & $\overline{K} N$ &$\pi \Lambda$ &$\pi \Sigma$ &$\eta \Sigma$ &$\overline{K} \Delta$ &$\pi \Sigma^*$
\\ \cline{2-7}
PW     &D          &D           &D         &D          &D        &D \\
LO  &39(3)      &27(3)       &3.0(1.2)     &0.08(0.01)      &1.6(0.2)    &7(1)  \\
NLO  & 55(12)       &14(4) & 0.6(0.8)      & 0.22(0.06)      & 3.9(0.8)     &7.4(2.3)  \\
Exp      &48(7)      &20.4(4.4)   &4.2(2)    &           &         &12(2.8) \\
\hline \hline
\end{tabular}
\label{ngttab10}
\end{center}
\end{table}
\vspace*{1cm}
\begin{table}[!h]
\caption{The decay widths of $\Xi$ states in ${\bf 70}$-plet whose
mass is currently experimentally known. Values are in MeV.}
\begin{center}
\begin{small}
\begin{tabular}{l ccc cc }
\hline \hline
&\multicolumn{5}{c}{$\Xi(1820)$}
\\
\cline{2-6}
 &\multicolumn{2}{c}{$\pi \Xi ^*$} &$\overline{K} \Lambda$ &$\overline{K} \Sigma$ &$\pi \Xi$ \\
\cline{2-6}
PW     &S          &D           &D         &D          &D     \\
LO  & 2.3(0.6)    & 2.6(0.3)        &10(1)     &14(1)    &4.2(0.9) \\
NLO  &2.4(2.2)       & 3.2(0.6)      & 18(3)     & 29(4)       &0.3(0.6) \\
Exp      &           &            &          &           &      \\
\hline \hline
\end{tabular}
\label{ngttab12}
\end{small}
\end{center}
\end{table}
\begin{table}[!h]
\caption{The decay widths of $\Delta$ states in ${\bf 70}$-plet
whose mass is currently experimentally known. Values are in MeV.}
\begin{center}
\begin{tabular}{l ccc cccc }
\hline \hline
&\multicolumn{2}{c}{$\Delta(1620)$} & &\multicolumn{4}{c}{$\Delta(1700)$}
\\
\cline{2-3}\cline{5-8}
 &$\pi N$ &$\pi \Delta$ &&\multicolumn{2}{c}{$\pi \Delta$} &$\pi N$ &$K \Sigma$  \\
\cline{2-3}\cline{5-8}
PW     &S          &D          &&S          &D     &D          &D         \\
LO  &34(5)      &62(7)     && 215(39)     & 20(4) &22(4)   &$\sim 0$    \\
NLO  &34(12)    &64(14)     &&157(52)       &18(8)  & 18(11)       &$\leq 0.04$     \\
Exp      &35.7(7.4)  &64.3(21.7) &&112(53)    &12(10)&45(21)     &          \\
\hline \hline
\end{tabular}
\label{ngttab11}
\end{center}
\end{table}
\newpage

\section{Summary and conclusions}
This work has extended previous analyses based on the $1/N_c$ expansion of the low lying negative parity baryon partial decay widths. The extension includes all known decay channels, and was carried out to first sub-leading order in the $1/N_c$ expansion and first order in $SU(3)$ symmetry breaking. The approximations involved were the following ones: for $SU(3)$ preserving amplitudes only up to 2-body operators and for $SU(3)$ breaking amplitudes only 1-body operators were included.  Mixings between states in different $SU(3)$ multiplets as well as $SU(6)\times O(3)$ configuration mixings  were neglected. These approximations are necessary due to the limitations in the available empirical inputs for the partial decay widths.
One important focus of the analysis was on $SU(3)$ symmetry breaking, whose significance is noticed first by   violations in LO  coefficient independent relations. In fact, the symmetry breaking effects are of unnaturally large magnitude as the NLO analysis shows. On the other hand, the $1/N_c$ expansion seems to work rather well, as the coefficients of NLO order symmetry preserving operators are in general of natural or smaller than natural magnitude. This agrees with the conclusions drawn in previous work where the non-strange sector had been analyzed.  The existence of other solutions to the NLO fit where the $\chi^2$ is not very significantly larger than the one presented in this work is an issue. Such solutions do have however unnaturally large NLO coefficients and lead to an inconsistent $1/N_c$ expansion.
Obviously,  additional and more accurate partial decay widths should be available for settling this issue.
An interesting open problem left by the analysis are the two S-wave decays $N(1700)\to \pi \Delta$ and $\Lambda(1690)\to \pi \Sigma^*$, which are predicted to be rather large, and are very correlated in the analysis. More accurate information on these widths has the potential to modify the results for the S-wave coefficients and the mixing angle $\theta_{N_1}$ in turn.

From the present work one can draw some conclusions that seem quite robust, namely, that the 1-body operator plays a key role in both S- and D-wave decays, as one would have expected from the quark  model picture, and that 2-body operators are crucial for obtaining an overall consistent description,  although they have smaller strength than the 1-body ones. Some uncertainties concerning mixing angles, in particular those affected by large $SU(3)$ breaking corrections seem to remain.
Further refinement of the present analysis, with the aim at improving in particular the determination of mixing angles, can be carried out by simultaneously analyzing  masses, decays and photo-couplings of the {\bf 70}-plet baryons \cite{nextpaper}. In addition, present progress in the analysis of recent and new data  will very likely lead to an improvement on the inputs  allowing  for a refinement of the present analysis and conclusions.

\section*{ACKNOWLEDGEMENTS}
We thank Michael D\"oring, Hiroyuki Kamano, Victor Mokeev and Igor Strakovsky for informative discussions. JLG  and NNS   thank  the Grupo de Particulas y Campos, Centro At\'omico Bariloche, and in particular Professor Roberto Trinchero, for the  hospitality
extended to them during completion of part of this work. 
This work was supported by DOE Contract No. DE-AC05-06OR23177 under which JSA operates the Thomas Jefferson
National Accelerator Facility,  by the National Science Foundation (USA) through grant   PHY-0555559 and   PHY-0855789 (JLG and CJ),  by CONICET (Argentina)  grant \# PIP 00682 and by ANPCyT  (Argentina)
grants  PICT 07-03-00818 (NNS).

\newpage
\section{Appendix: $SU(3)$ Isoscalar Factors}

This appendix gives the   isoscalar factors needed for the calculations carried out in this work. They correspond to the emission of mesons
belonging to an {\bf 8}  of $SU(3)$, and are given for the irreducible representations of $SU(3)$ of interest for generic $N_c$.
We denote  the isoscalar factors by:
\beq
\left(\begin{array}{cc}
     (p,q)   & (1,1)       \\
     Y~I    & y ~i
\end{array}  \right\Arrowvert \left.     \begin{array}{c}
      (p',q')      \\
     Y'~~ I'
\end{array} \right)_\gamma\, ,
\eeq
where the $SU(3)$
representations are identified in terms of the two labels defining the Young tableu, namely  $(p,q)$, where $p+2 q=N_c$.

For baryons, the  correspondences between multiplets for generic odd $N_c$ and $N_c=3$ are as follows:
$(p=0, q=\frac{N_c-3}{2}) \rightarrow   {\bf 1}$,  $(p=1, q=\frac{N_c-1}{2}) \rightarrow   {\bf 8}$,
 and
$(p=3, q=\frac{N_c-3}{2})\rightarrow   {\bf 10}$.
Table \ref{tabXV} displays these correspondences more explicitly.

\begin{center}
\begin{table}[h]
\caption{Representation correspondences for arbitrary odd $N_c$.  Displayed are the $SU(3)$ generic $N_c$ multiplets corresponding to  the ones at $N_c=3$, namely  {\bf 1}, {\bf 8} and {\bf 10}.  }
\vspace{.5cm}
\begin{tabular}{ccccccccccc}
\hline \hline
\multicolumn{2}{c}{{\bf 1} Baryons} &\hspace{.5cm} \ &
\multicolumn{2}{c}{{\bf 8} Baryons} & \hspace{.5cm}\ &
\multicolumn{2}{c}{{\bf 10} Baryons} & \hspace{.5cm}\ &
\multicolumn{2}{c}{Mesons} \\
\cline{1-2}\cline{4-5}\cline{7-8}\cline{10-11}
\multicolumn{2}{c}{$(p,q) = (0,\frac{N_c-3}{2})$}& \hspace{.5cm}\ &
\multicolumn{2}{c}{$(p,q) = (1,\frac{N_c-1}{2})$}&\hspace{.5cm} \ &
\multicolumn{2}{c}{$(p,q) = (3,\frac{N_c-3}{2})$}&\hspace{.5cm} \ &
\multicolumn{2}{c}{$(p,q) = (1,1)$} \\[2mm]
\cline{1-2}\cline{4-5}\cline{7-8}\cline{10-11}
State &  $(Y,I)$ & & State &  $(Y,I)$ & & State  & $(Y,I)$  & & State  & $(Y,I)$ \\ [1mm]
\cline{1-2}\cline{4-5}\cline{7-8}\cline{10-11} &&&&& & & &
\\[-3.02mm]
$\Lambda$ & $(\frac{N_c-3}{3},0)$ && N & $(\frac{N_c}{3},\frac{1}{2})$ && $\Delta$  & $(\frac{N_c}{3},\frac{3}{2})$ && $\pi$ &  $(0,1)$ \\[2mm]
&&&$\Sigma$&  $(\frac{N_c-3}{3},1)$ &&  $\Sigma^*$  &$(\frac{N_c-3}{3},1)$  && $\eta$ & $(0,0)$ \\[2mm]
&&&$\Lambda$&  $(\frac{N_c-3}{3},0)$  && $\Xi^*$  &$(\frac{N_c-6}{3},\frac{1}{2})$  && K & $(1,\frac12)$ \\[2mm]
&&&$\Xi$&  $(\frac{N_c-6}{3},\frac{1}{2})$ && $\Omega$  &$(\frac{N_c-9}{3},0)$  && $\overline{K}$  & $(-1,\frac12)$ \\[2mm]
\hline \hline
\end{tabular}
\\[4mm]
\label{tabXV}
\end{table}
\end{center}

Table XVI through XX give the isoscalar factors of  interest. The first row of the tables depict the excited baryon state, and the second row the final baryon and meson. The isoscalar factor arguments are in the order $\left( B^*, P\mid\mid B\right)$, where $B^*$, $P$ and $B$ are the quantum numbers of the excited baryon, the meson and the final baryon respectively.

\begin{center}
\begin{table}
\parbox{5in}{\caption{Isoscalar factors for  ${\bf 8} \to {\bf 8}$ decays.
The listed
values should be multiplied by $f_1 = \frac{1}{(N_c+3)}$ and
$f_2 = \frac{1}{(N_c+3)}\sqrt{\frac{(N_c-1)}{(N_c+7)}}$
to obtain the actual isoscalar factors for $\gamma=1$ and $\gamma=2$, respectively.}}\\
\vspace{.3cm}
\begin{tabular}{cc cccc c cccc}
\hline \hline
& &\multicolumn{4}{c}{$N$} &
  &\multicolumn{4}{c}{$\Lambda$}
\\ \cline{3-6}\cline{8-11}
& & $  \eta N$ & $\pi N  $ & $K \Sigma  $ & $K \Lambda $ &
  & $ \overline{K}N$ & $\pi \Sigma $ & $ \eta \Lambda $ & $ K \Xi  $
\\ \cline{3-6}\cline{8-11}
&&&&&&
\\[-3mm]
$\gamma=1$ & & $N_c$ & 3 & $\sqrt{3(N_c-1)}$ & $\sqrt{3(N_c+3)}$ &
          & $-\sqrt{\frac{3(N_c+3)}{2}}$ & 0 & $N_c-3$ &
                       3$\sqrt{\frac{N_c-1}{2}}$
\\[2mm]
$\gamma=2$ & & 3 & $-(N_c+6)$ & $\frac{N_c+15}{\sqrt{3(N_c-1)}}$ & $- \sqrt{3(N_c+3)}$ &
          & $\sqrt{\frac{3(N_c+3)}{2}}$ &  $-\sqrt{\frac{(N_c+3)^3}{3(N_c-1)}}$ & 6 &
               $\frac{9-N_c}{\sqrt{2(N_c-1)}}$
\\[3mm]
\hline
\end{tabular}
\\[3mm]
\begin{tabular}{cc cc cc cc}
\hline \hline
      & &\multicolumn{5}{c}{$\Sigma$}
\\ \cline{3-7}
       & & $ \overline{K}N$ & $\eta \Sigma $ & $\pi \Sigma $ & $\pi \Lambda $ & $K\Xi  $
\\
\cline{3-7} &&&&&&
\\[-3mm]
$\gamma=1$ & & $3\sqrt{\frac{N_c-1}{2}}$ & $N_c-3$ & $2 \sqrt{6}$ &
          0 & $\sqrt{\frac{3(N_c+3)}{2}}$
\\[2mm]
$\gamma=2$ & & $\frac{N_c+15}{\sqrt{2(N_c-1)}}$ &
    $\frac{2(N_c-9)}{N_c-1}$ & $-\sqrt{\frac{2}{3}} \frac{(N_c-3)(N_c+7)}{N_c-1}$ &
       $\sqrt{\frac{(N_c+3)^3}{N_c-1}}$ & $-\frac{5N_c+3}{N_c-1}
       \sqrt{\frac{N_c+3}{6}}$
\\[3mm] \hline
\end{tabular}
\\[3mm]
\begin{tabular}{cc cc cc}
\hline \hline & &\multicolumn{4}{c}{$\Xi$}
\\
\cline{3-6} & & $ \overline{K}\Sigma$ & $ \overline{K} \Lambda $ & $\eta  \Xi $ & $\pi \Xi $
\\
\cline{3-6} &&&&&
\\[-3mm]
$\gamma=1$ & & $-\sqrt{N_c+3}$ & 3$\sqrt{N_c-1}$ & $N_c-6$ & 3
\\[2mm]
$\gamma=2$ & & $\frac{5N_c+3}{3(N_c-1)} \sqrt{N_c+3}$ &
$\frac{9-N_c}{\sqrt{N_c-1}}$ & $\frac{7N_c-15}{N_c-1}$ &
$\frac{N_c^2 + 3 N_c +36}{3(N_c-1)}$
\\[2mm] \hline
\end{tabular}
\end{table}
\end{center}

%\pagebreak
%%%%%%%%%%%%%%%%%%%
\begin{center}
\begin{table}
\parbox{5in}{\caption{Isoscalar factors for   ${\bf 10} \to {\bf 10}$  decays. The listed
values should be multiplied by $f_1 = \frac{1}{\sqrt{45+N_c(N_c+6)}}$ and
$f_2 = \sqrt{\frac{5(N_c-3)(N_c+5)}{(N_c+1)(N_c+9)(45+N_c(N_c+6))}}$
to obtain the actual isoscalar factors for $\gamma=1$ and $\gamma=2$, respectively.  Note that $f_2$ vanishes at $N_c=3$, as in that case there is a unique recoupling ${\bf 10}\otimes {\bf 8}\to {\bf 10}$.
}}\\ \vspace{.3cm}
\begin{tabular}{cc ccc c cccc}
\hline \hline
& & \multicolumn{3}{c}{$\Delta$} &
  & \multicolumn{4}{c}{$\Sigma^*$}
\\
\cline{3-5}\cline{7-10}
& & $\eta \Delta  $ & $\pi \Delta $ & $K \Sigma^* $ &
  & $ \overline{K} \Delta $ & $\eta \Sigma^* $ & $\pi \Sigma^* $ & $K \Xi^* $
\\
\cline{3-5}\cline{7-10}
&&&&& &&&&
\\[-3mm]
$\gamma=1$& & $N_c$ & $3 \sqrt{5}$ & $\sqrt{3(N_c+5)}$ &
         & $-\frac{3 \sqrt{N_c+5}}{2}$ & $N_c-3$ & $2 \sqrt{6}$ & $\sqrt{6(N_c+3)}$
\\[2mm]
$\gamma=2$& & 3 & $-\frac{N_c+6}{\sqrt{5}}$ & $\frac{3-N_c}{\sqrt{3(N_c+5)}}$ &
         & $\frac{N_c-3}{2 \sqrt{N_c+5}} $ & $\frac{4(N_c+3)}{N_c+5}$ &
                   $-\frac{N_c^2 + 10 N_c + 33}{\sqrt{6}
                  (N_c+5)}$ & $\frac{3-N_c}{N_c+5} \sqrt{\frac{2(N_c+3)}{3}}$
\\[2mm] \hline
\end{tabular}
\\[3mm]
%%%%%%%%%%%%%
\begin{tabular}{c c cc c c ccc}
\hline \hline
& &\multicolumn{4}{c}{$\Xi^*$} &
  &\multicolumn{2}{c}{$\Omega$}
\\
\cline{3-6} \cline{8-9}
& & $ \overline{K} \Sigma^* $ & $\eta\Xi^* $ & $\pi \Xi^* $ & $ K \Omega $ &
  & $  \overline{K} \Xi^* $ & $ \eta \Omega $
\\
\cline{3-6} \cline{8-9} & & & & &
\\[-3mm]
$\gamma=1$& &$-2 \sqrt{N_c+3}$ & $N_c-6$ & 3 & $3 \sqrt{N_c+1}$ &
         & $-3 \sqrt{\frac{N_c+1}{2}}$ & $N_c-9$
\\[2mm]
$\gamma=2$& & $\frac{2(N_c-3) \sqrt{N_c+3}}{3 (N_c+5)} $ &
           $\frac{5N_c+9}{N_c+5}$ & $-\frac{N_c^2+ 9 N_c+36}{3(N_c+5)}$ &
           $\frac{(3-N_c) \sqrt{N_c+1}}{N_c+5}$ &
         & $\frac{N_c-3}{N_c+5} \sqrt{\frac{N_c+1}{2}}$ & $\frac{6(N_c+1)}{N_c+5}$
\\[2mm] \hline
\\
\end{tabular}
\end{table}
\end{center}

\begin{center}
\begin{table}
\parbox{5in}{\caption{Isoscalar factors for    ${\bf 8} \to {\bf 10}$ decays. The listed
values should be multiplied by
$f= \frac{\sqrt{2}}{\sqrt{(N_c+1)(N_c+5)}}$ to obtain
the actual isoscalar factors.}}\\
\vspace{.3cm}
\begin{tabular}{ccc ccc c}
\hline \hline
& &\multicolumn{2}{c}{$N$}&
  &\multicolumn{2}{c}{$\Lambda$}
\\
\cline{3-4}\cline{6-7}
& & $\ \ \ \ \ \  \pi \Delta  \ \ \ \ \ \ $ & $\ \ \ \ \ \  K \Sigma^*  \ \ \ \ \ \ $ &
  & $\ \ \ \ \ \ \ \ \ \pi  \Sigma^* \ \ \ \ \ \ \ \ \ $ & $\ \ \ \ \ \ K  \Xi^* \ \ \ \ \ \ $
\\
\cline{3-4}\cline{6-7}
&&& &&&
\\[-3mm]
& & $-\sqrt{\frac{(N_c-1)(N_c+5)}{2}}$ & $-2\sqrt{\frac{N_c-1}{3}}$ &
  & $-\sqrt{\frac{(N_c+3)(N_c-1)}{3}}$ &  $-\sqrt{2 (N_c-1)}$
\\[2mm] \hline
\end{tabular}
\\[3mm]
 
\begin{tabular}{cc cccc c cccc}
\hline \hline
& &\multicolumn{4}{c}{$\Sigma$} &
  & \multicolumn{4}{c}{$\Xi$}
\\
\cline{3-6} \cline{8-11}
& & $\ \ \   \overline{K}\Delta  \ \ \ $ & $\ \ \  \eta  \Sigma^*  \ \ \ $ & $\ \ \  \pi \Sigma^*   \ \ \ $ & $\ \ \ K   \Xi^*  \ \ \ $ &
  & $\ \ \ \eta   \Xi^* \ \ \ $ & $\ \ \ \pi  \Xi^* \ \ \ $ & $\ \ \  \overline{K} \Sigma^*  \ \ \ $ & $\ \ \   K  \Omega \ \ \ $
\\
\cline{3-6} \cline{8-11}
&&&&& &&&&&
\\[-3mm]
& & $\sqrt{N_c+5}$ & 2 & $\frac{N_c+1}{\sqrt{6}}$ & $\sqrt{\frac{2(N_c+3)}{3}}$ &
  & 2 & $\frac{2 N_c}{3}$ & $\frac{2\sqrt{N_c+3}}{3}$ & $2 \sqrt{N_c+1}$
\\[2mm] \hline
\\
\end{tabular}
\end{table}
\end{center}

\begin{center}
\begin{table}
\parbox{5in}{\caption{Isoscalar factors for    ${\bf 10} \to {\bf 8}$ decays. The listed
values should be multiplied by
$f=\left( (N_c+7)(N_c-1) \right)^{-1/2}$
to obtain the actual isoscalar factors.}}\\
\vspace{.3cm}
\begin{tabular}{cc cc cc cc cc }
\hline \hline & &\multicolumn{2}{c}{$\Delta$}& & &
\multicolumn{4}{c}{$\Xi^*$}
\\
\cline{3-4}\cline{7-10}
 & & $ \pi N $ & $K\Sigma $ & & & $\ \ \ \   \overline{K}  \Sigma \ \ \ \ $ & $\ \   \overline{K}  \Lambda   \ \ $ & $\ \  \eta \Xi   \ \ $ & $\ \ \pi  \Xi \ \ $
\\
\cline{3-4}\cline{7-10} & & & & & & & &
\\[-3mm]
& & $-\sqrt{(N_c-1)(N_c+5)}$ & $2 \sqrt{\frac{N_c+5}{3}}$ & &
  & $\frac{2 \sqrt{N_c+3}}{3}$ & $2 \sqrt{N_c-1}$ & $-2$ & $-\frac{2N_c}{3}$
\\[2mm] \hline
\end{tabular}
\\[3mm]
 
\begin{tabular}{cc cc cc cc cc }
\hline \hline & &\multicolumn{5}{c}{$\Sigma^*$}& & & {$\Omega$}
\\
\cline{3-7}\cline{10-10}
 & & $\  \eta \Sigma  \ $ & $  \overline{K} N   $ & $\ \ \ \ \pi  \Sigma   \ \ \ \ $ & $\pi\Lambda  $ & $K \Xi $ & & & $ \overline{K}\Xi  $
\\
\cline{3-7}\cline{10-10}& & & & & & & & &
\\[-3mm]
 & & $-2$ & $\sqrt{2(N_c-1)}$ & $-\frac{N_c+1}{\sqrt{6}}$ & $-\sqrt{(N_c+3)(N_c-1)}$ & $\sqrt{\frac{2(N_c+3)}{3}}$
& & & $\sqrt{2(N_c+1)}$
\\[2mm] \hline
\\
\end{tabular}
\end{table}
\end{center}

\begin{center}
\begin{table}
\parbox{5in}{\caption{Isoscalar factors for    ${\bf 1} \to {\bf 8}$ decays.  }}\\
\vspace{.3cm}
\begin{tabular}{ cccc}
\hline \hline

\multicolumn{4}{c}{$\Lambda$}
\\ \cline{1-4}

   $ \overline{K}N $ & $ \pi \Sigma $ & $ \eta \Lambda  $ & $ K \Xi  $
\\ \cline{1-4}
 
\\[-3mm]

           $1$ &$ \sqrt{\frac{2}{N_c-1}}$ & $\sqrt{\frac{6}{N_c+3}}$ &
                       $\sqrt{\frac{12}{(N_c+3)(N_c-1)}}$
 \\[3mm]
\hline
\end{tabular}
\end{table}
\end{center}


\newpage

 
 
\begin{thebibliography}{99}
\bibitem{CR-QM}
S.~Capstick and W.~Roberts,
Prog.~Part.~Nucl.~Phys.~{\bf 45}, S241 (2000) and references therein.
 \bibitem{DM} R.~Dashen and A.~V.~Manohar,
Phys.~Lett.~{\bf B315}, 425 (1993);
Phys.~Lett.~{\bf B315}, 438 (1993).
\bibitem{CGKM}Ch.~D.~Carone, H.~Georgi, L.~Kaplan and D.~Morin,  Phys.~Rev.~{\bf D50},  5793 (1994).
\bibitem{Goity} J.~L.~Goity,  Phys.~Lett.~{\bf B414}, 140 (1997). 
\bibitem{PY}D.~Pirjol and T-M.~Yan, Phys.~Rev.~{\bf D57}, 5434 (1998);  Phys.~Rev.~{\bf D57}, 1449 (1998).
 \bibitem{GS} J.~L.~Gervais and B.~Sakita,
Phys.~Rev.~Lett.~{\bf 52}, 87 (1984);
Phys.~Rev.~{\bf D30}, 1795 (1984).
 \bibitem{CCGL}C.~E.~Carlson, Ch.~D.~Carone, J.~L.~Goity and  R.~F.~Lebed,  Phys.~Lett.~{\bf B438},  327 (1998);
  Phys.~Rev.~{\bf D59}, 114008 (1999).  
  \bibitem{SGS} C.~L.~Schat, J.~L.~Goity and N.~N.~Scoccola,  Phys.~Rev.~Lett.~{\bf 88}, 102002 (2002). \\
J.~L.~Goity, C.~L.~Schat and  N.~N.~Scoccola, Phys.~Rev.~{\bf D66}, 114014 (2002).
\bibitem{CL} T.~D.~Cohen and R.~F.~Lebed,  Phys.~Rev.~{\bf D67},  096008 (2003);  Phys.~Rev.~{\bf D68}, 056003 (2003).\\
T.~D.~Cohen, C.~Dakin, A.~Nellore  and R.~F.~Lebed,  Phys.~Rev.~{\bf D70},  056004 (2004).
\bibitem{GSS70decays} J.~L.~Goity, C.~L.~Schat  and N.~N.~Scoccola, Phys.~Rev.~{\bf D71}, 034016 (2005).
\bibitem{FKR} R.~P.~Feynman, M.~Kislinger and F.~Ravndal,
 Phys.~Rev.~{\bf D3},  2706 (1971).
 \bibitem{PDG}
Particle Data Group, C.~Amsler et al., J.~of Phys {\bf  G 37}, 075021 (2010).
\bibitem{thesis} C.~Jayalath, \lq\lq Baryon resonances and their decays in the $1/N_c$ expansion\rq\rq PhD thesis, Hampton University (2010), unpublished.
\bibitem{CL2} T.~D.~Cohen and R.~F.~Lebed, Phys.~Rev.~{\bf D70}, 096015 (2004).
\bibitem{Mixings} J.~L.~Goity, Phys.~Atom.~Nucl.~{\bf 68}, 624 (2005).
\bibitem{GdeUS} E.~Gonzalez de Urreta and N.~N.~Scoccola, in preparation.
\bibitem{Edmonds} A.~R.~Edmonds, \lq\lq Angular Momentum in Quantum Mechanics\rq\rq, Princeton University Press (1974).
\bibitem{Hecht} K.~T.~Hecht, Nucl.~Phys.~{\bf 62}, 1 (1965).
\bibitem{GMS} N.~N.~Scoccola, J.~L.~Goity, and N.~Matagne, Phys.~Lett.~{\bf B663},  222 (2008).
\bibitem{JGS}  J.~L.~Goity, C.~Jayalath and N.~N.~Scoccola, Phys.~Rev.~{\bf D80}, 074027 (2009). 
\bibitem{DJM} R.~F.~Dashen, E.~E.~Jenkins and A.~V.~Manohar,  Phys.~Rev.~{\bf D51}, 3697 (1995).
\bibitem{PirjolSchat}
D.~Pirjol and C.~Schat,
Phys.~Rev.~{\bf D67}, 096009 (2003).\\
T.~D.~Cohen and R.~F.~Lebed, Phys.~Rev.~Lett.~{\bf 91}, 012001 (2003).\\
T.~D.~Cohen and R.~F.~Lebed, Phys.~Rev.~{\bf D72}, 056001 (2005).
\bibitem{nextpaper} J.~L.~Goity, E.~Gonzalez de Urreta, C.~Jayalath and N.~N.~Scoccola, work in progress.
\bibitem{GW} R.~A.~Arndt et al, Phys.~Rev.~{\bf C74},  045205 (2006).
\bibitem{Juelich}M.~D\"oring et al, Nucl.~Phys.~{\bf A829}, 170 (2009).
\bibitem{EBAC} N.~Suzuki et al, Phys.~Rev.~Lett.~{\bf 104},  042302 (2010).
\bibitem{Manley-Thoma} D.~M.~Manley and E.~M.~Saleski, Phys.~Rev.~{\bf D45}, 4002 (1992).\\
U.~Thoma et al., Phys.~Lett.~{\bf B659}, 87 (2008).
\bibitem{Mokeev} V.~Mokeev, private communication.
\end{thebibliography}
\end{document}